\documentclass[print]{aa}
\usepackage[varg]{txfonts}
\usepackage{physics}
\usepackage{subcaption}
\usepackage{textcomp}

\usepackage{array}
\usepackage{booktabs}
\usepackage{tikz}

\newcommand{\comment}[1]{}

\newcommand{\red}[1]{\textcolor{black}{#1}}
\begin{document}




\title{Large Interferometer For Exoplanets (LIFE):}

\titlerunning{LIFE: IV. Array architectures for a space-based MIR nulling interferometer}

\authorrunning{Hansen et al.}

\subtitle{IV. Ideal kernel-nulling array architectures for a space-based mid-infrared nulling interferometer}

\author{Jonah T. Hansen\inst{1}\thanks{Correspondence: \href{mailto:jonah.hansen@anu.edu.au}{jonah.hansen@anu.edu.au}},
  Michael J. Ireland\inst{1}\and the LIFE Collaboration\inst{2}}

\institute{Research School of Astronomy and Astrophysics, College of Science, Australian National University, Canberra, Australia, 2611
\and 
\url{www.life-space-mission.com}}

\date{Received X XXX XXXX}

\abstract {} {Optical interferometry from space for the purpose of detecting and characterising exoplanets is seeing a revival, specifically from missions such as the proposed Large Interferometer For Exoplanets (LIFE). A default assumption since the design studies of \textit{Darwin} and TPF-I has been that the Emma X-array configuration is the optimal architecture for this goal. Here, we examine whether new advances in the field of nulling interferometry, such as the concept of kernel-nulling, challenge this assumption.} {We develop a tool designed to derive the photon-limited signal to noise ratio of a large sample of simulated planets for different architecture configurations and beam combination schemes. We simulate four basic configurations: the double Bracewell/X-array, and kernel-nullers with three, four and five telescopes respectively.} {We find that a configuration of five telescopes in a pentagonal shape, using a five aperture kernel-nulling scheme, outperforms the X-array design in both search (finding more planets) and characterisation (obtaining better signal, faster) when total collecting area is conserved. This is especially the case when trying to detect Earth twins (temperate, rocky planets in the habitable zone), showing a 23\% yield increase over the X-array. On average, we find that a five telescope design receives 1.2 times the signal over the X-array design.} {With the results of this simulation, we conclude that the Emma X-array configuration may not be the best architecture choice for the upcoming LIFE mission, and that a five telescope design utilising kernel-nulling concepts will likely provide better scientific return for the same collecting area, provided that technical solutions for the required achromatic phase shifts can be implemented.}

\maketitle 

\section{Introduction}

Optical interferometry from space remains one of the key technologies that promises to bring an unprecedented look into high angular resolution astrophysics. This is particularly true in the booming field of exoplanet detection and characterisation; the Voyage 2050 plan of the European Space Agency (ESA) \citep{2021ESAVoyage} recently recommended that the study of temperate exoplanets and their atmospheres in the mid-infrared (MIR) be considered for a large scale mission, given it can be proven to be technologically feasible. The Large Interferometer For Exoplanets (LIFE) initiative \citep{LIFEPaper1} was developed to achieve this goal - using a space-based MIR nulling interferometer to find and characterise temperate exoplanets around stars that would be otherwise too challenging or unfeasible for other techniques such as single aperture coronography or transit spectroscopy.

The notion of using a space-based nulling interferometer to look for planets is not new. It was first proposed by \cite{1978Bracewell}, and then further developed by \cite{1995Leger} and \cite{1997Angel} among others, leading to two simultaneous concept studies of large scale missions by ESA (the \textit{Darwin} mission) and NASA (the Terrestrial Planet Finder - Interferometer (TPF-I) mission). Unfortunately to those excited by the prospect of a space-based interferometer, both missions were dropped by their corresponding agencies due to a combination of funding issues, technical challenges and lack of scientific understanding of the underlying exoplanet population. 

These concerns have been tackled substantially in the recent decade - exoplanet space missions such as \textit{Kepler} \citep{Kepler}, TESS \citep{TESS} and CHEOPS \citep{CHEOPS}, and radial velocity surveys on instruments including HIRES \citep{HIRES} and HARPS \citep{HARPS}, have provided the community with a vast trove of knowledge concerning exoplanet demographics  \citep[e.g.,][]{2018Petigura,2018Fulton,2020Berger,2021HansenPlanets}. On the technical side, missions such as PROBA-3 \citep{2018Loreggia} and Pyxis \citep[Hansen et al. in prep]{2020HansenHonours} aim to demonstrate formation-flight control at the level required for interferometry, while the Nulling Interferometry Cryogenic Experiment (NICE) \citep{2020Gheorghe} aims to demonstrate MIR nulling interferometry at the sensitivity required for a space mission under cryogenic conditions. Other developments such as in MIR photonics \citep{2017Harry,2019Gretzinger} and MIR detectors \citep{2020Cabrera} have also progressed to the point where a space-based MIR interferometer is significantly less technically challenging.

In this light, we wish to revisit the studies into array architecture that were conducted during the \textit{Darwin}/TPF-I era and identify which architecture design (with how many telescopes) is best suited to both detect and characterise exoplanets. Since the initial trade-off studies \citep{2005LayXarray}, the assumed architecture of an optical/MIR nulling interferometer has been the Emma X-array configuration - four collecting telescopes in a rectangle formation that reflect light to an out-of-plane beam combiner. Functionally, this design acts as two Bracewell interferometers with a $\pi/2$ phase chop between them. However, new developments in the theory of nulling interferometry beam combination, particularly that of kernel-nulling \citep{2018Martinache,2020Laugier}, \red{allow other configurations to obtain the same robustness as the X-array, and so} may no longer render this the best configuration for the purposes of the proposed LIFE mission.

\red{Kernel-nulling is a generalised concept that allows the beam combination of multiple telescopes to be robust to piston and optical path delay (OPD) errors to second-order. The X-array described above, with the $\pm\, \pi/2$ phase chop (sine-chop) between the two nulled arms, is equivalent to a kernel nuller that is signal-to-noise optimal at a single wavelength; offering similar robustness against piston errors \citep{2003Velusamy,2004LayNoise}. The advantages of kernel-nullers (or equivalently the sine-chopped X-array) also extend to the removal of symmetric background sources such as local zodiacal light and exozodiacal light \citep{2010Defrere} due to their asymmetric responses. This symmetric source suppression was the reason that the sine-chop was favoured over the cosine-chop (chopping between 0 and $\pi$) in the original investigations of the double Bracewell \citep{2003Velusamy}. These benefits lead us to postulate that all competitive nullers for exoplanet detection must be kernel-nullers, or alternatively offer the same benefits even if they are not labelled as such (e.g. the double Bracewell/X-array).}

Here, we build a simulator that identifies the signal to noise ratio (SNR) of a simulated population of planets for a given telescope array architecture, to identify which configuration is best suited for both the detection of an exoplanet (where the orbital position of a planet may not be known) and characterisation of an exoplanet's atmosphere (where the orbital position of a planet is known). Four architectures are compared: the default \red{sine-chopped} X-array configuration and three kernel-nullers with three, four and five telescopes respectively. \red{To compare these configurations, we base our analysis on the relative merits of the transmission maps associated with each architecture.  We highlight here that we only consider the photon noise limited case, with instrumental errors discussed in a follow up paper \citep{LIFE7}.}

\section{Model Implementation}

In our model, we adapt a similar spectral range to that of the initial LIFE study \citep[][hereafter labelled LIFE1]{LIFEPaper1}, spanning 4 to 19~\textmu m. We assume $N_\lambda$ = 50 spectral channels, giving a spectral bandpass of 0.3~\textmu m per channel. 

\subsection{Star and Planet Populations}

To draw our population of planets, we use the P-Pop simulator tool\footnote{GitHub: \url{https://github.com/kammerje/P-pop}}, as found in \cite{2018Kammerer}. This tool uses Monte-Carlo Markov Chains to draw a random population of planets with varying orbital and physical parameters around a set input catalogue of stars. We chose to use the LIFE star catalogue (version 3), as described in LIFE1. This catalogue contains main sequence stars primarily of spectral types F through M within 20\,pc without close binaries; histograms of the stellar properties in the input catalogue can be found in Fig. A.1 of LIFE1. The underlying planet population was drawn using results from NASA's ExoPaG SAG13 \citep{2018Kopparapu}, with planetary radii spanning between 0.5 and 14.3~R$_\oplus$ (the lower and upper bounds of atmosphere-retaining planets covered in \cite{2018Kopparapu}) and periods between 0.5 and 500\,days. As in LIFE1, binaries with a separation greater than 50~AU are treated as single stars, and all planets are assumed to have circular orbits. We run the P-Pop simulator 10 times to produce 10 universes of potential planetary parameters, in order for our calculation of SNR detection rates to be robust. We also highlight to the reader that the underlying population used in this paper is different to the one used in LIFE1, and that comparisons between the planet detection yields in these works must be made with caution.

To create our stellar and planetary photometry, we approximate the star as a blackbody radiator and integrate the Planck function over the spectral channels. For the planet, we consider two contributions: thermal and reflected radiation. The thermal radiation is generated by approximating the planet as a blackbody, and utilising the effective temperature and radius generated from P-Pop. For the effective temperature, we note that P-pop uses a random number for the Bond albedo in the range $A_\text{B} \in [0,0.8)$. The reflected radiation on the other hand is the host star's blackbody spectrum scaled by the semi-major axis and albedo of the planet. That is, we assume the reflected spectral radiance is related to the host star by
\begin{equation}
    B_\text{planet, ref}(\lambda) = A_\text{g} \cdot f_p \cdot\left(\frac{R_p}{a}\right)^2B_\text{star}(\lambda),
\end{equation}
where $A_\text{g}$ is the mid-infrared geometric albedo from P-pop ($A_\text{g}\in[0,0.1)$), $f_p$ is the Lambertian reflectance, $R_p$ is the radius of the planet and $a$ is the semi-major axis. We note here, however, that typically the reflected light component of an exoplanet's flux is negligible when compared to the thermal radiation in the MIR wavelength regime. 

\subsection{Architectures}
\label{Sec:architectures}
\red{In this work, we define two specific modes for the array - search and characterisation. The search mode is where the array is optimised for a single, predefined radius around a star, nominally the habitable zone, with the aim of detecting new planets. The array spins, and any modulated signal can then be detected as a planet. It should be noted that the planet is almost certainly not in the optimised location of the array. For our purposes, \red{we consider ideal signal extraction, with a single planet only}. We also \red{assume the array rotates an integer number of symmetry angles (e.g. 72 degrees for the 5-telescope array)} during an observation. For the characterisation mode, we assume we know the angular position of the planet we are characterising and optimise the array for that position. As such, no array rotation is required. This stage focuses on obtaining the highest signal of the planet possible} 

We consider four architectures to compare in this analysis: the X-array design based on the double Bracewell configuration (the default choice inherited from the Darwin/TPF-I trade studies \citep{2005LayXarray}), and then three sets of kernel-nullers based on the work of \cite{2018Martinache} and \cite{2020Laugier}, using three, four and five telescopes respectively. We assume that the beam combining spacecraft is out of the plane of the collector spacecraft (the `Emma' configuration) so that any two-dimensional geometry can be realised.

To obtain the responses of each architecture, we create a transmission map of each telescope configuration normalised by the flux per telescope. We first must define a reference baseline to optimise the array around - we have chosen to use the shortest baseline in any of the configurations (that is, adjacent telescopes). If we wish to maximise the response of the interferometer at a given angular separation from the nulled centre $\delta$, the baseline should be calculated as
\begin{equation}
    B = \Gamma_B\frac{\lambda_B}{\delta},
\end{equation}
where $\lambda_B$ is a reference wavelength (that is, a wavelength for which the array is optimised) and $\Gamma_B$ is \red{an architecture dependent} scale factor. This factor must be configured for each different architecture as well as for each nulled output. That is, the array may be optimised for only a single nulled output.

The intensity response on-sky is given by
\begin{align}
    \vb{R}_l(\vb*{\alpha}) = \left|\sum_{k} \vb{M}_{k,l} e^{2\pi i\vb{u}_k\vdot\vb*{\alpha}}\right|^2,
\end{align}
where $\vb{M}_{k,l}$ is the transfer matrix of the beam combiner, taking $k$ telescopes and turning them into $l$ outputs\red{; $\vb{u}_k$ are the locations of the telescopes in units of wavelength; and $\vb*{\alpha}$ is the angular on-sky coordinate.}

The response of different architectures are thus defined by their telescope coordinates \red{$\vb{u}_k$} and their transfer matrix $\vb{M}_{k,l}$. Each architecture will produce a different number of robust observables ($N_K$), \red{which are generated from linear combinations of the response maps $\vb{R}_l(\vb*{\alpha})$} \citep{2018Martinache}. Note that we have made the approximation that the electric field is represented by a scalar, and are not considering systematic instrumental effects. We have, however, provided some technical requirements on phase stability that will be discussed in Sect. \ref{Sec:results_search}, and will examine systematic instrumental effects in a follow-up publication \red{\citep{LIFE7}}.

\red{For the following discussion, all configuration diagrams and response maps can be found in Fig. \ref{Architectures}.}




\subsubsection{X-array}
We first consider the default assumed architecture of the LIFE mission (as described in LIFE1) and that of the \textit{Darwin} and TPF-I trade studies \citep{2004LayNoise,2005LayXarray} - the X-array or Double Bracewell. This design consists of four telescopes in a rectangle, with a shorter `nulling' baseline (defined here as $B$) and a longer `imaging' baseline (defined through the rectangle ratio $c$, such that the length is $cB$). The pairs of telescopes along each nulling baseline are combined with a $\pi$ phase shift along one of the inputs. Then, the two nulled outputs are combined with a $\pi/2$ phase chop. A diagram of the arrangement is in \red{Fig. \ref{Architectures}.A.a}.

We simply define the Cartesian positions of the telescopes as
\begin{align}
    x_k &= [0.5B,-0.5B,-0.5B,0.5B], \\
    y_k &= [0.5cB,0.5cB,-0.5cB,-0.5cB].
\end{align}
The transfer matrix of the system can be written as
\begin{equation}
    \vb{M}_{k,l} = \frac{1}{2}\begin{bmatrix}
    \sqrt{2} & \sqrt{2} & 0 & 0\\
    1 & -1 & i & -i\\
    i & -i & 1 & -1\\
    0 & 0 & \sqrt{2} & \sqrt{2},
\end{bmatrix}
\end{equation}
thus providing us with $k=4$ four outputs. We can see from the matrix that the responses $\vb{R}_0$ and $\vb{R}_3$ contain the majority of the starlight, and the other two phase chopped nulled outputs contain the planet signal. A single robust observable is simply given by the difference in intensities of these two outputs ($N_K = 1$). The observable, and its associated transmission map, are thus:
\begin{align}
    \vb{K}_1(\vb*{\alpha}) &=  \vb{R}_1(\vb*{\alpha}) - \vb{R}_2(\vb*{\alpha}) & \vb{T}_1(\vb*{\alpha}) = \red{\frac{1}{2}\left(\tilde{\vb{R}}_1(\vb*{\alpha})+\tilde{\vb{R}}_2(\vb*{\alpha})\right)}.
\end{align}
\red{We note here that in practice, when considering radially symmetric emission sources such as stellar or exozodiacal leakage, the two responses matrices are equivalent and so the transmission map can be functionally written as either $\tilde{\vb{R}}_1$ or $\tilde{\vb{R}}_2$. This isn't true for all orientations of edge on disks, and hence we have defined the transmission as the average of the two responses.}

\red{To identify the baseline scale factor $\Gamma_B$ that provides maximum sensitivity, we calculated the modulation efficiency of the array as a function of radial position in units of $\lambda/B$. The modulation efficiency $\xi$, as defined in \cite{LIFE2} and described in \cite{2004LayNoise} is a measurement of the signal response that a source should generate as the array is rotated. This is essentially a root-mean-squared (RMS) average over azimuthal angles, given as a function of radius by \cite{LIFE2}:
\begin{equation}
    \xi_i(r) = \sqrt{\langle\vb{K}_i(r,\phi)^2\rangle_\phi}.
\end{equation}
Note however that in this work we normalise the modulation efficiency by the flux per telescope rather than the total flux.}

\red{Assuming a rectangle ratio of $c=6$ (that is, imaging baseline is six times the size of the nulling baseline as suggested by \cite{2006Lay}), we plot the modulation efficiency as a function of radius in Fig. \ref{Architectures}.A.b. The position which gives a maximum efficiency indicates that the baseline scale factor should be $\Gamma_B = 0.59$. A map of the observable, in units of $\lambda_B/B$, is shown in Fig. \ref{Architectures}.A.c, along with a circle highlighting the radial separation that has the highest modulation efficiency.}

\begin{figure*}
    \centering
    \begin{tabular}{>{\centering\arraybackslash} m{1cm} >{\centering\arraybackslash} m{5cm} >{\centering\arraybackslash} m{5cm} >{\centering\arraybackslash} m{5cm} }
       \toprule
        Figure & A) X-array & B) Kernel-3 & C) Kernel-4 \\
        \midrule
        a & \includegraphics[width = 0.8\linewidth]{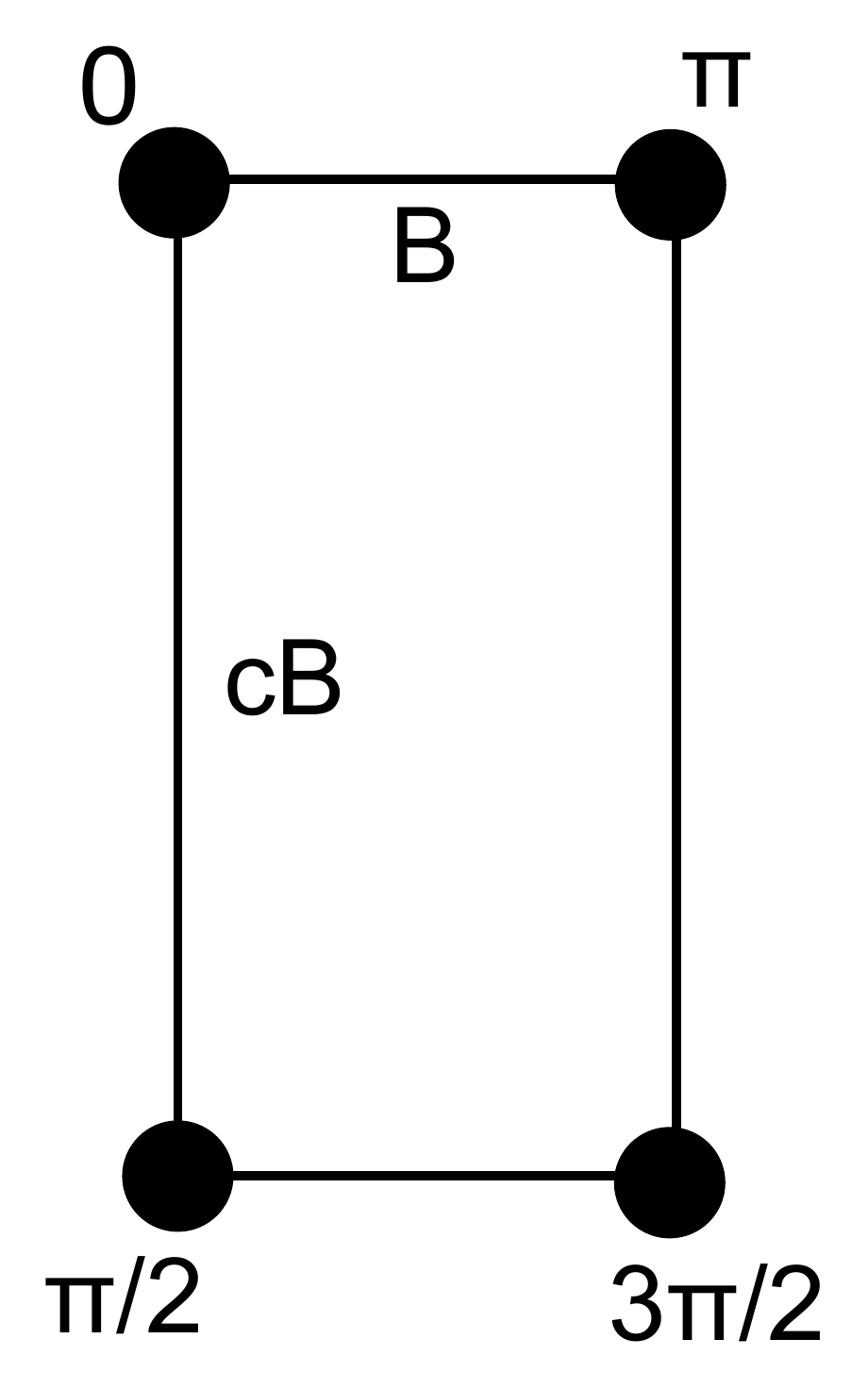} & \includegraphics[width = 0.8\linewidth]{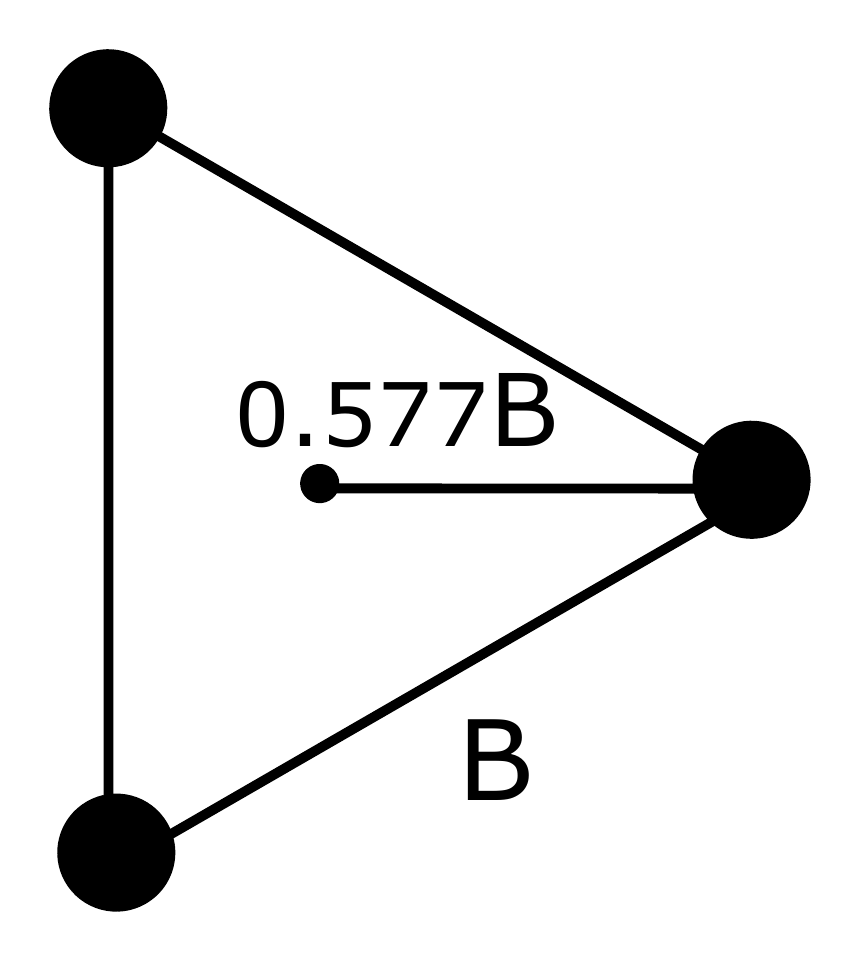} & \includegraphics[width = 0.8\linewidth]{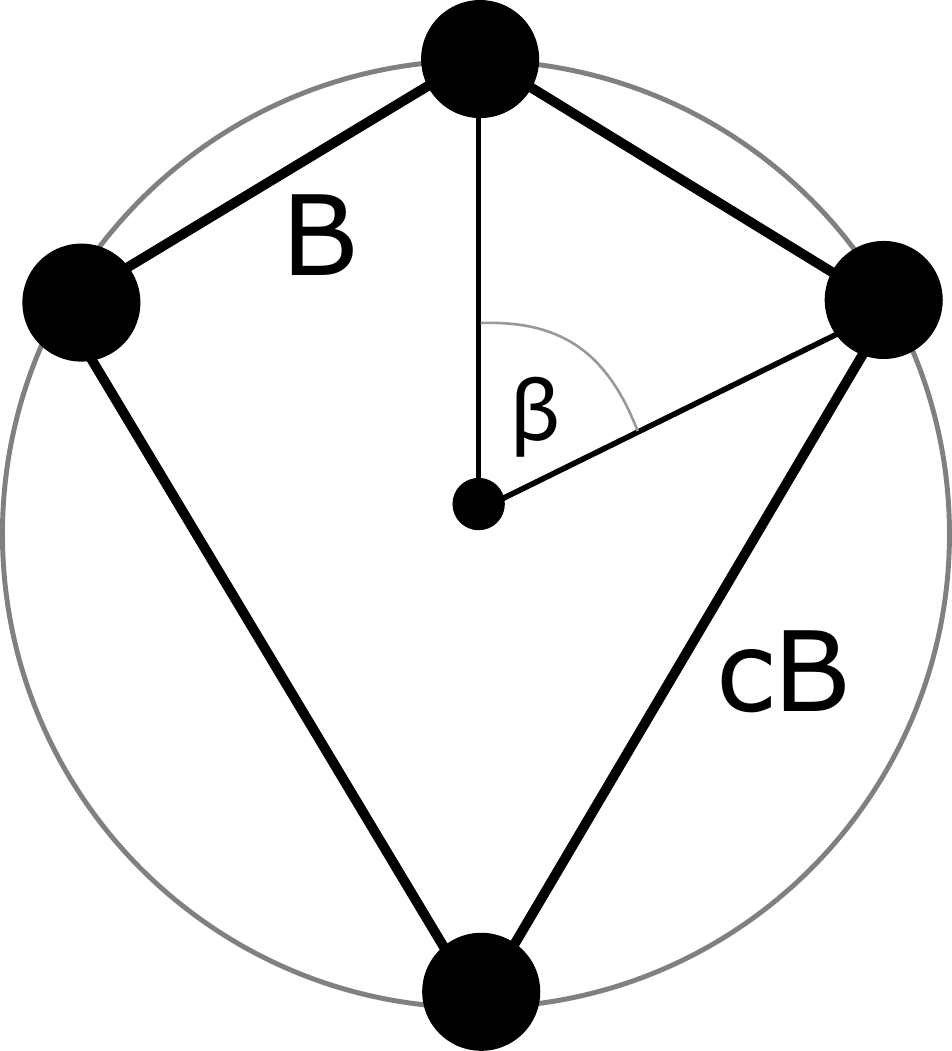} \\
        b & \includegraphics[width = \linewidth]{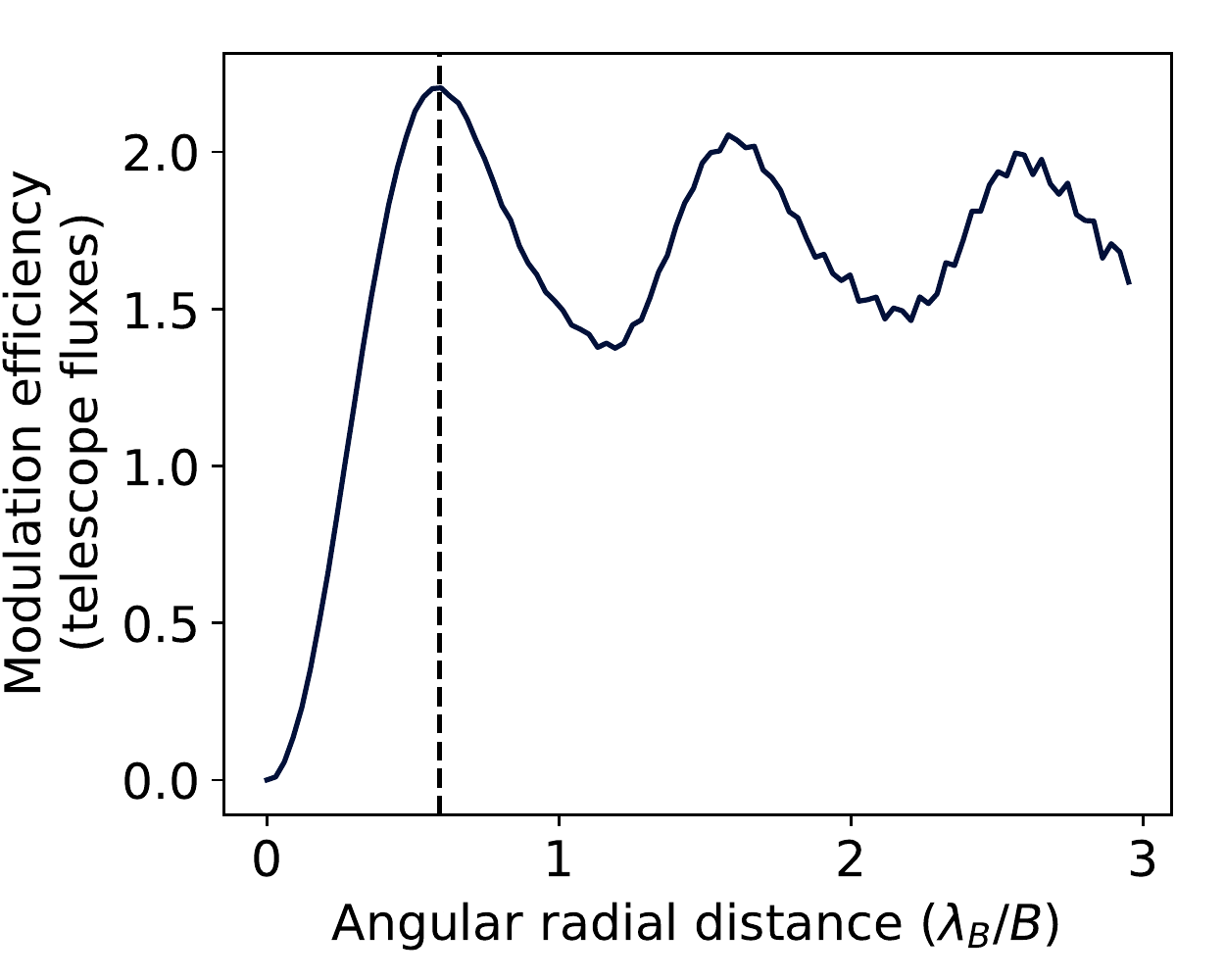} & \includegraphics[width = \linewidth]{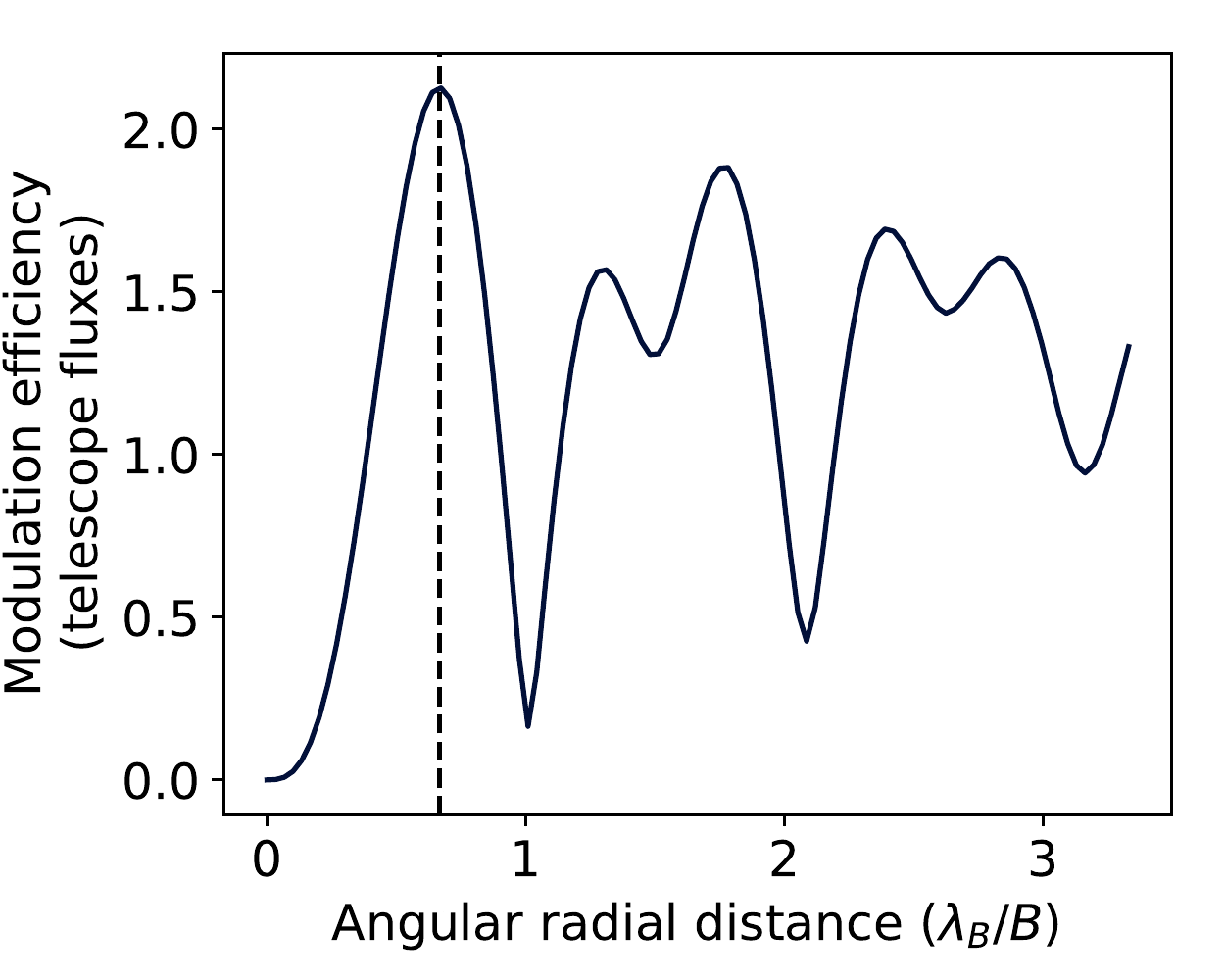} & \includegraphics[width = \linewidth]{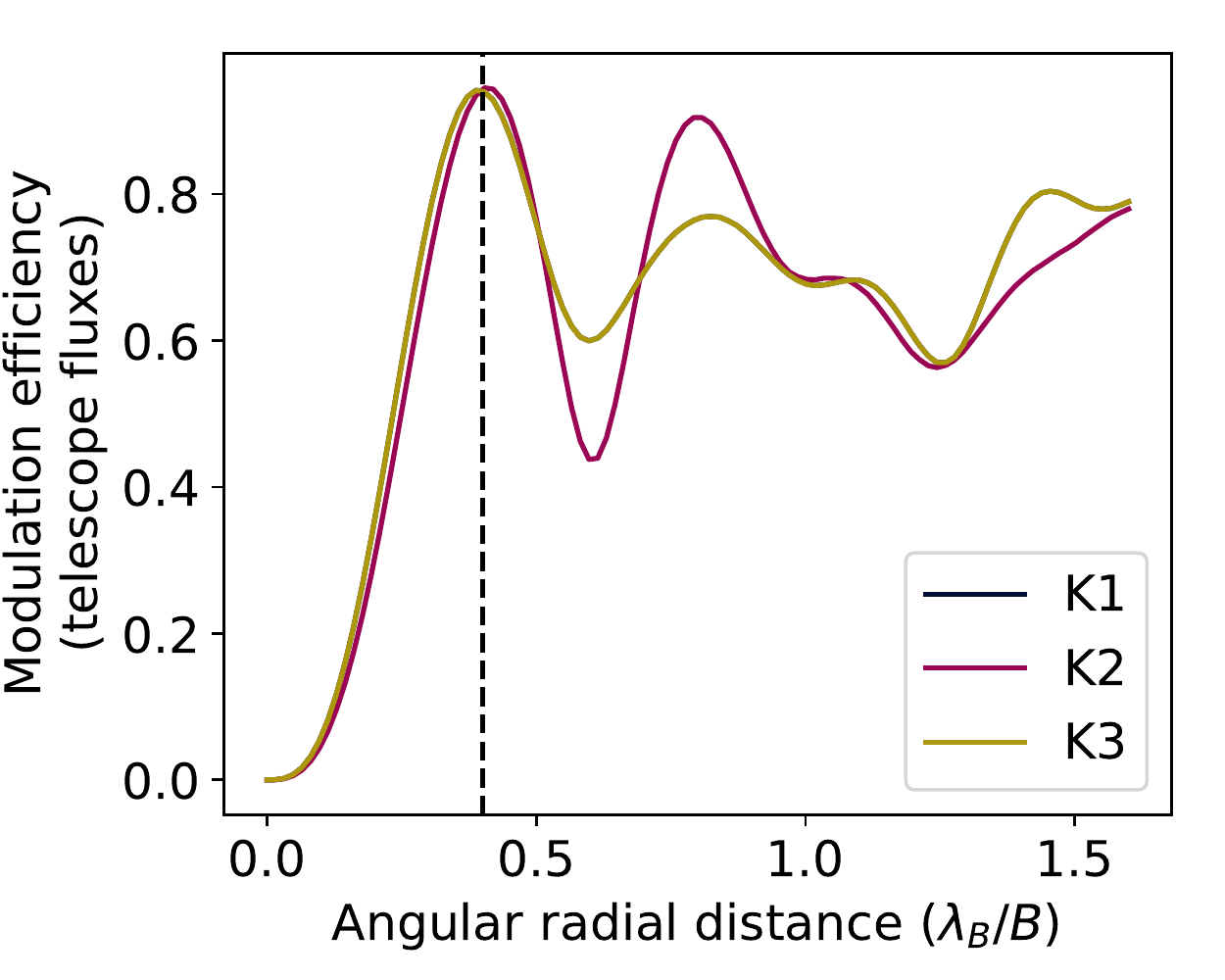} \\
         & & & \includegraphics[width = \linewidth]{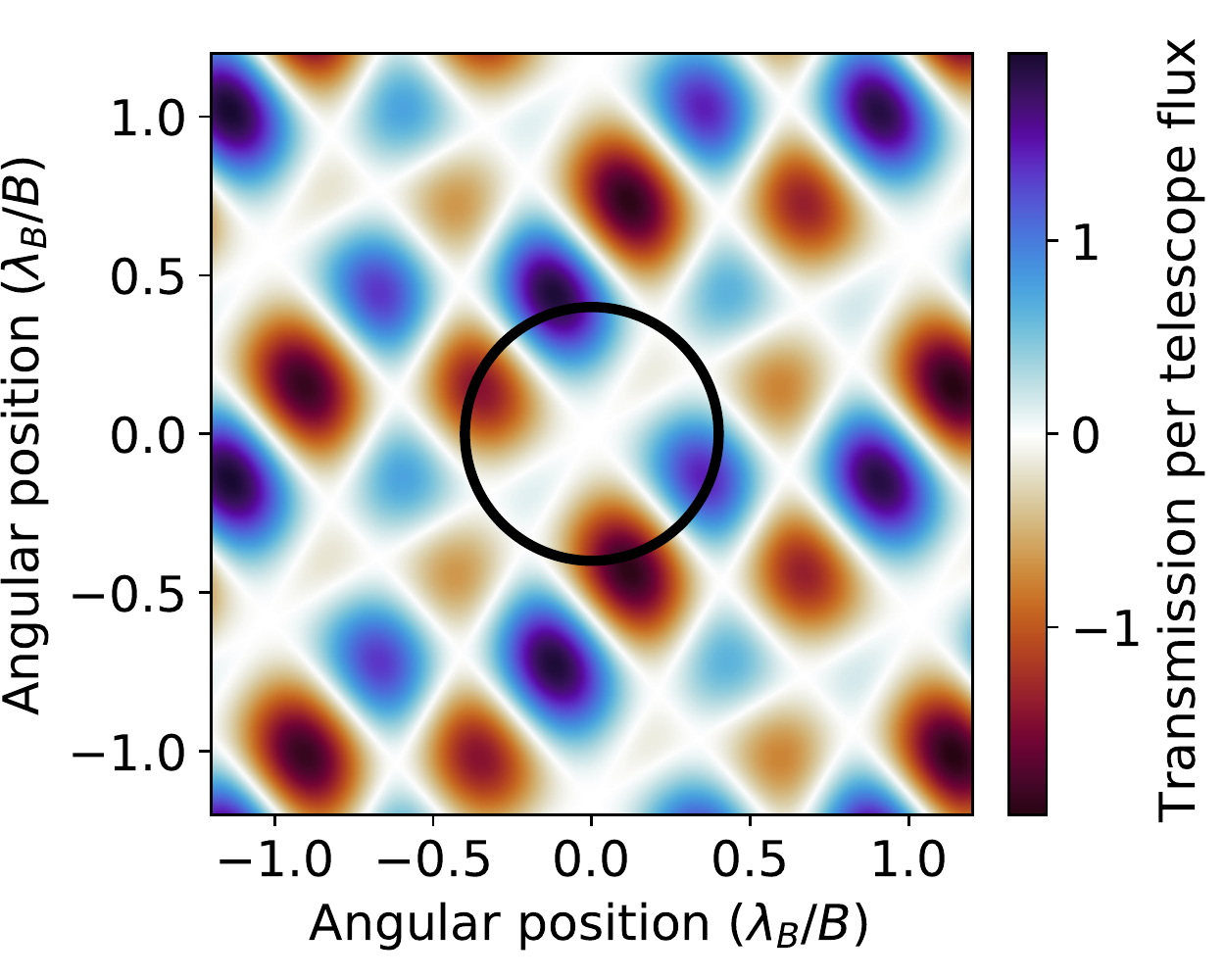} \\
         c & \includegraphics[width = \linewidth]{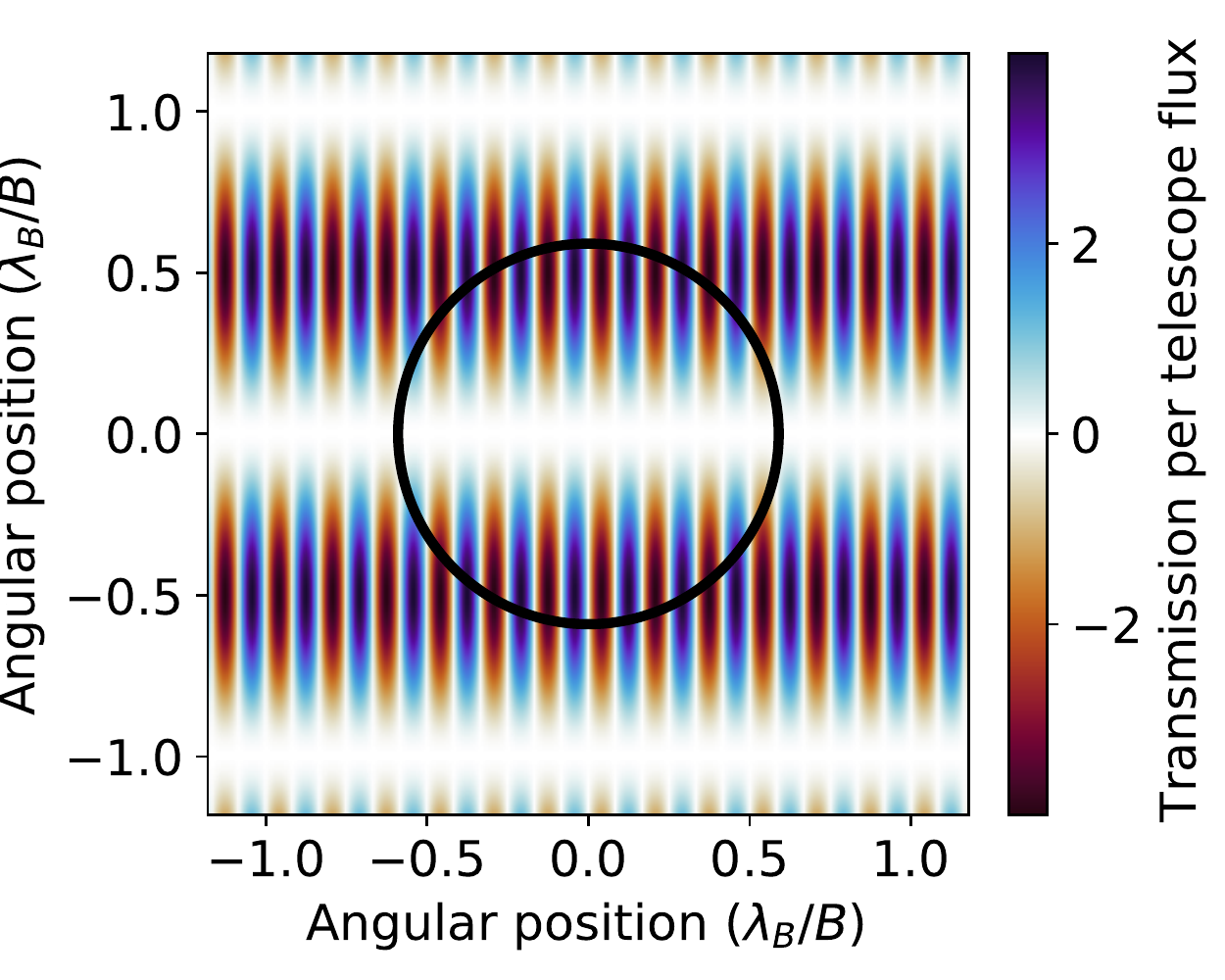}& \includegraphics[width = \linewidth]{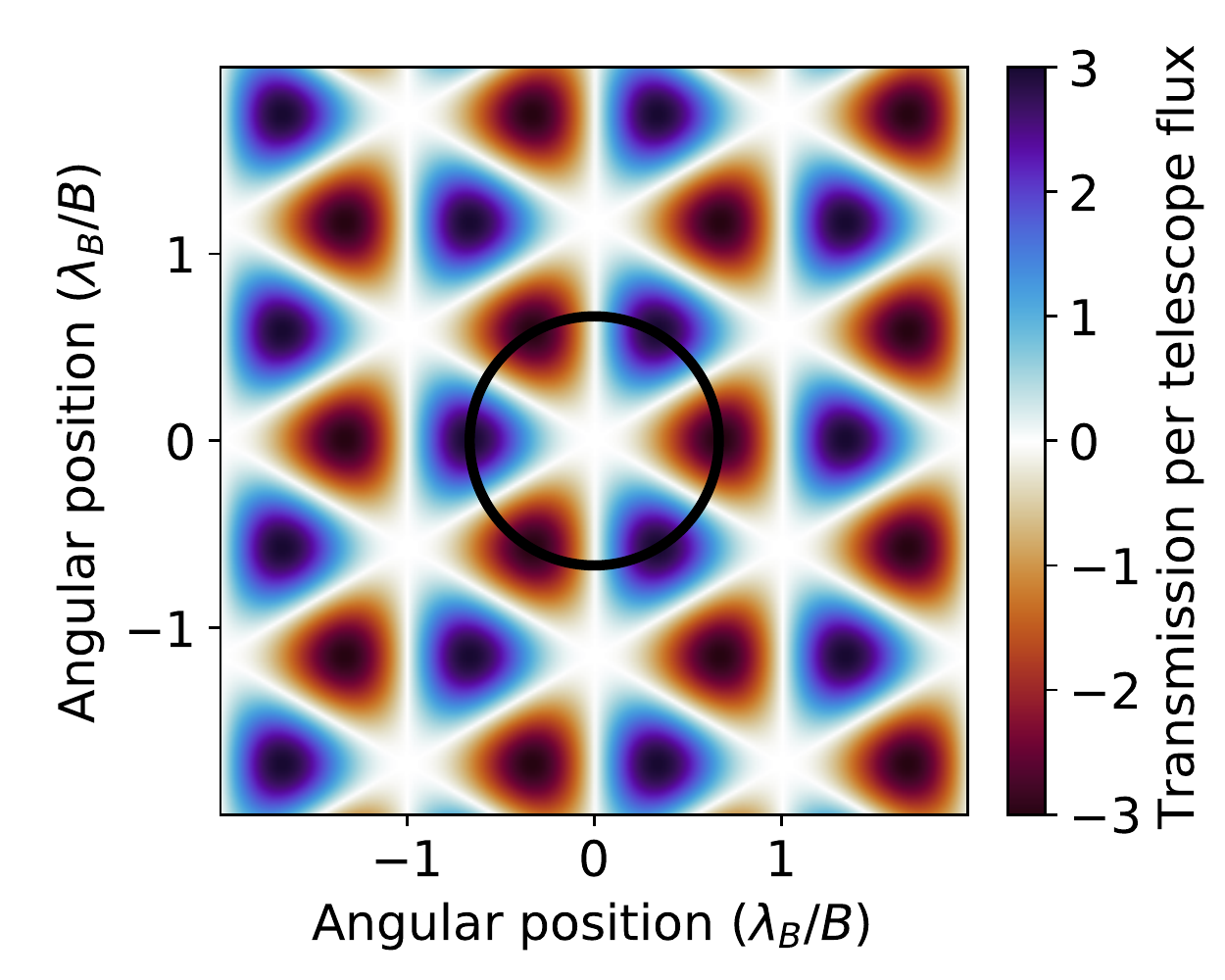} & \includegraphics[width = \linewidth]{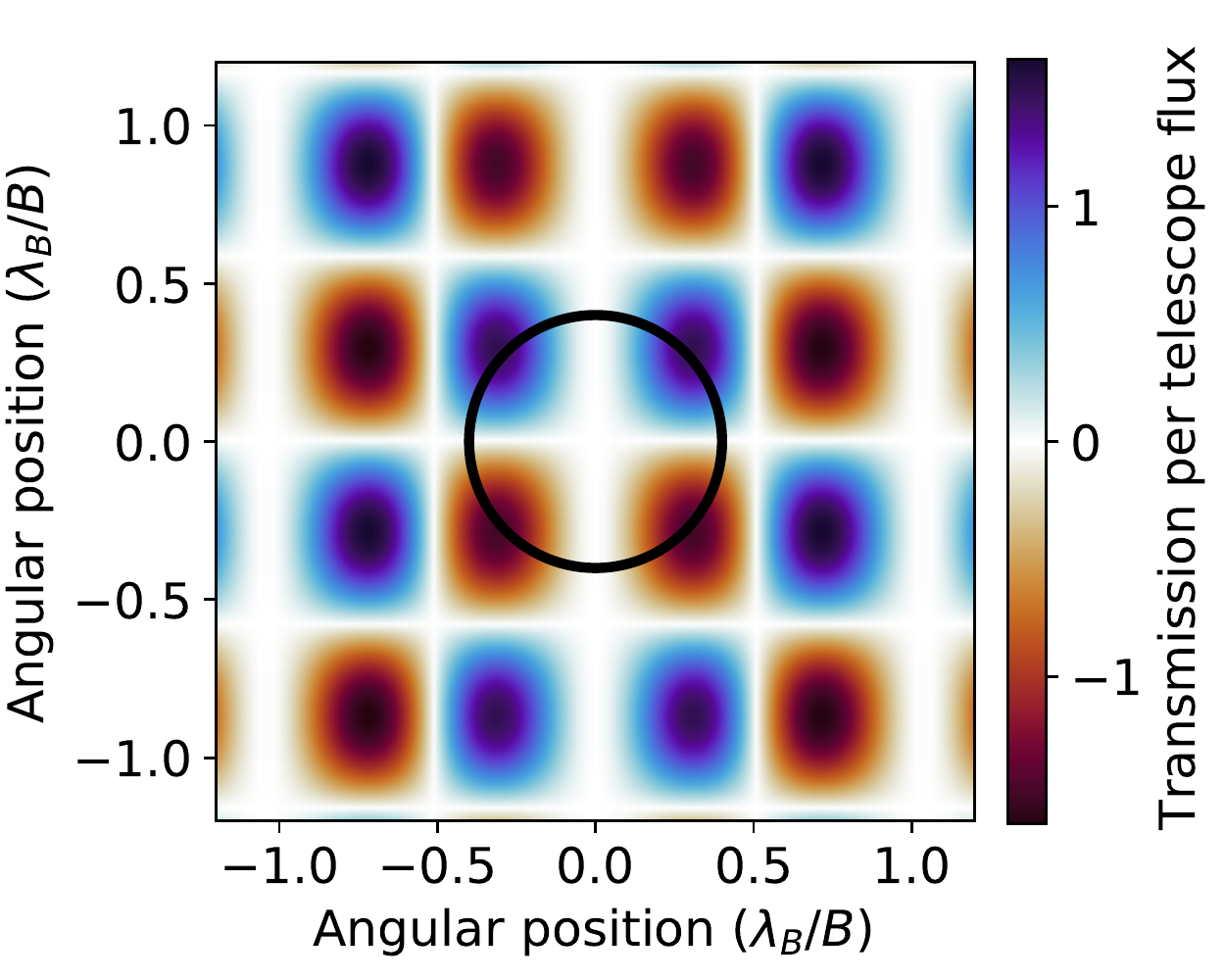} \\
         & & & \includegraphics[width = \linewidth]{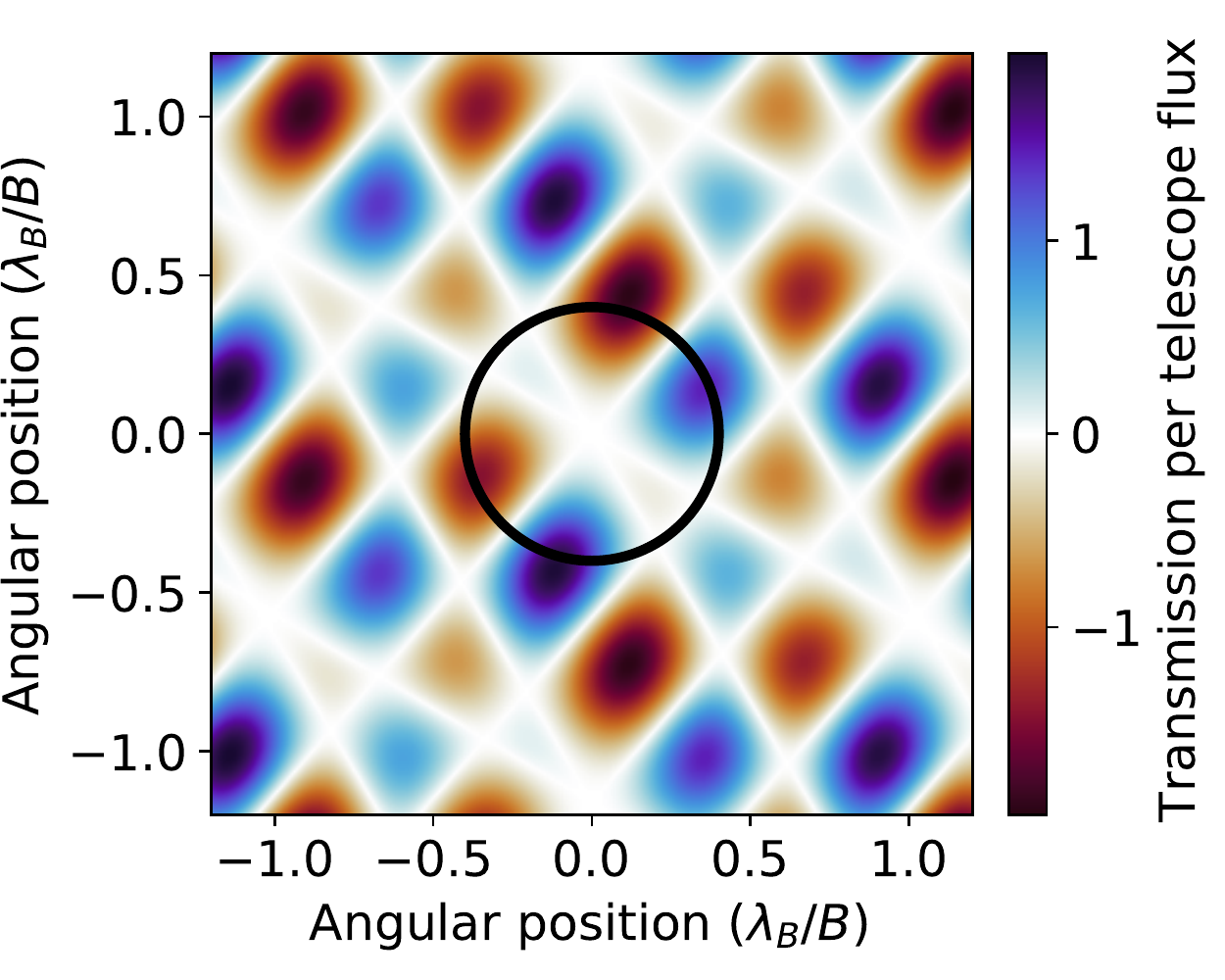} \\
        \bottomrule
    \end{tabular}
    \caption{\red{This figure is continued on the next page.}}
\end{figure*}
\begin{figure*}
    \centering
    \ContinuedFloat
    \begin{tabular}{>{\centering\arraybackslash} m{1cm} >{\centering\arraybackslash} m{5cm} >{\centering\arraybackslash} m{5cm} >{\centering\arraybackslash} m{5cm} }
       \toprule
        Figure & D) Kernel-5 (0.66) & E) Kernel-5 (1.03) & F) Kernel-5 (1.68) \\
        \midrule
        a & \includegraphics[width = 0.8\linewidth]{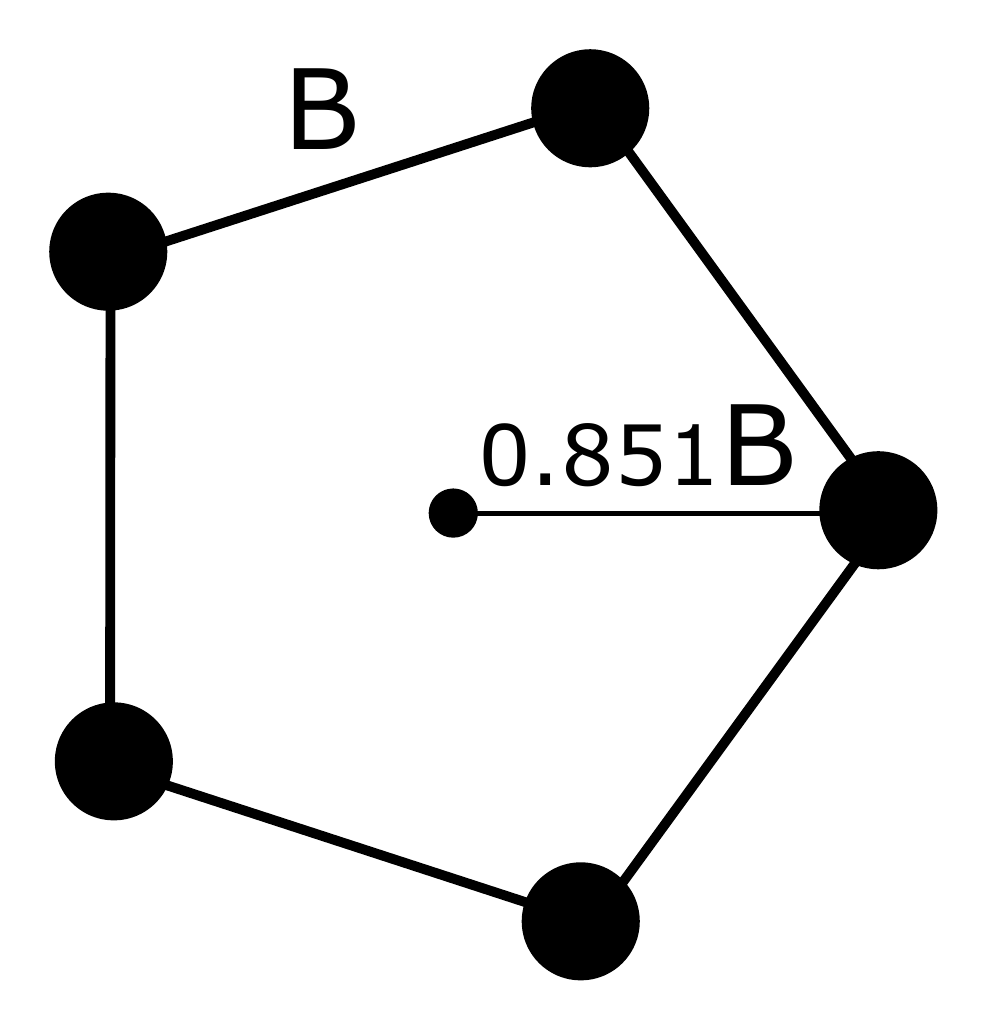} & \includegraphics[width = 0.8\linewidth]{images/5K_diagram.pdf} & \includegraphics[width = 0.8\linewidth]{images/5K_diagram.pdf} \\
        b & \includegraphics[width = \linewidth]{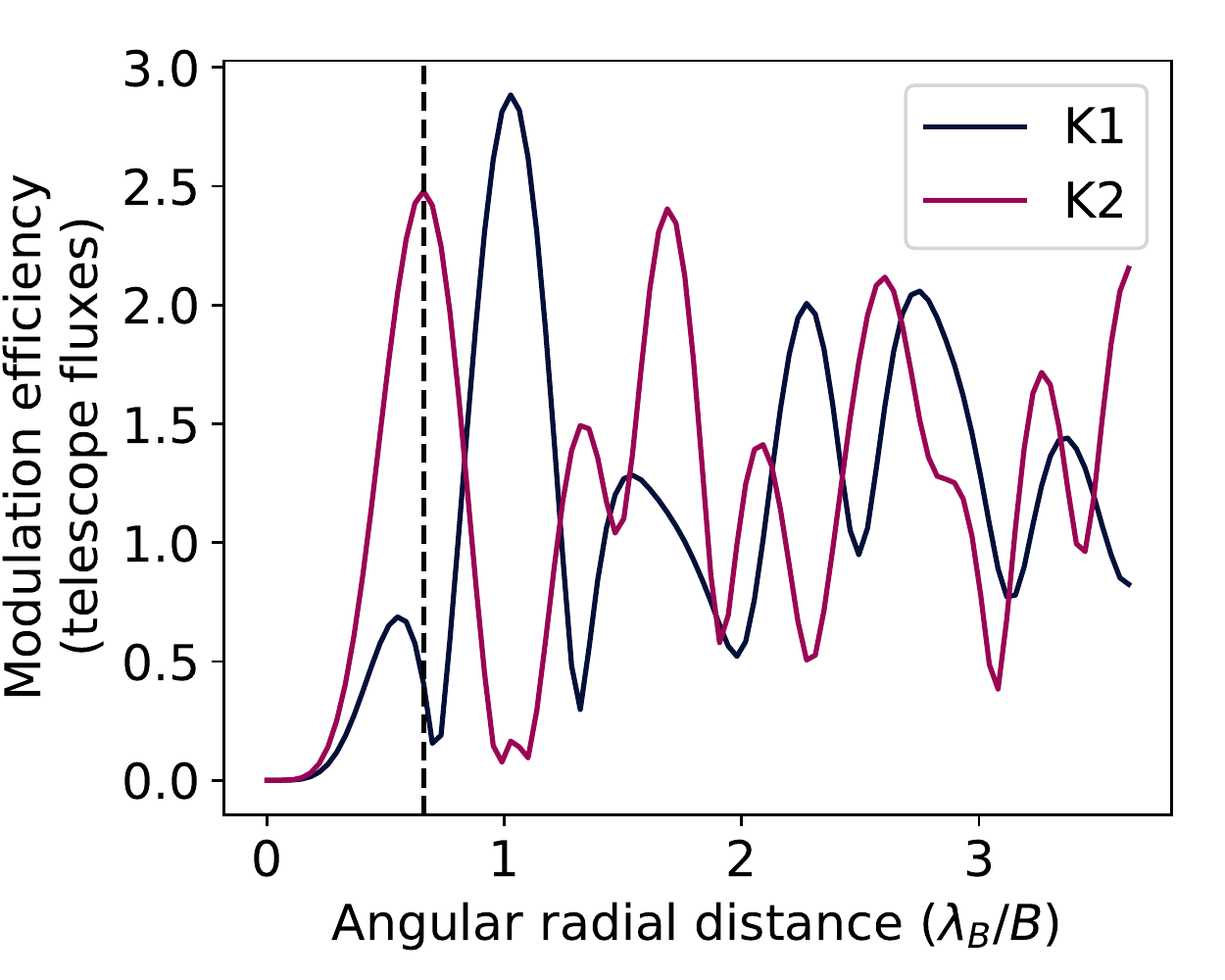} & \includegraphics[width = \linewidth]{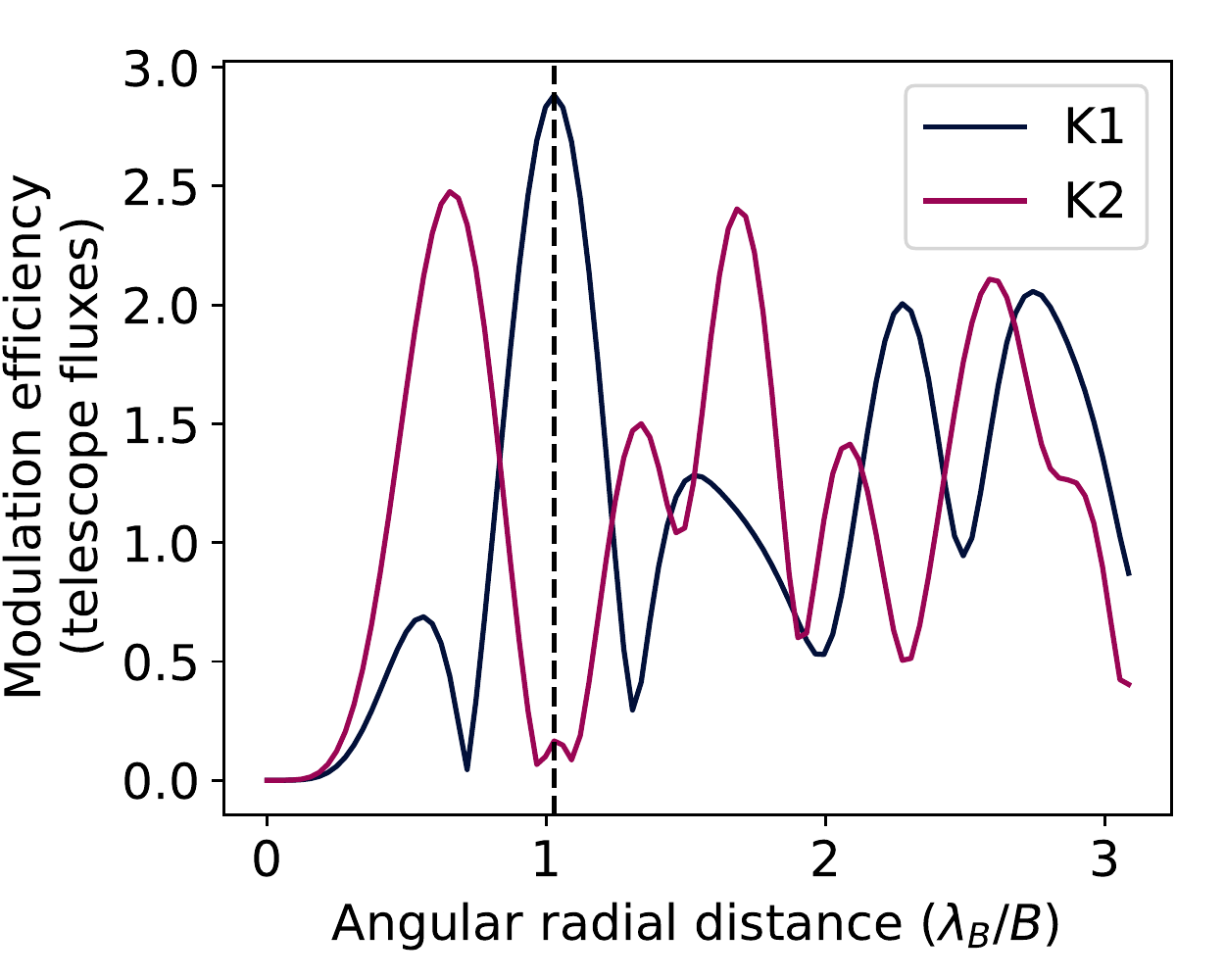} & \includegraphics[width = \linewidth]{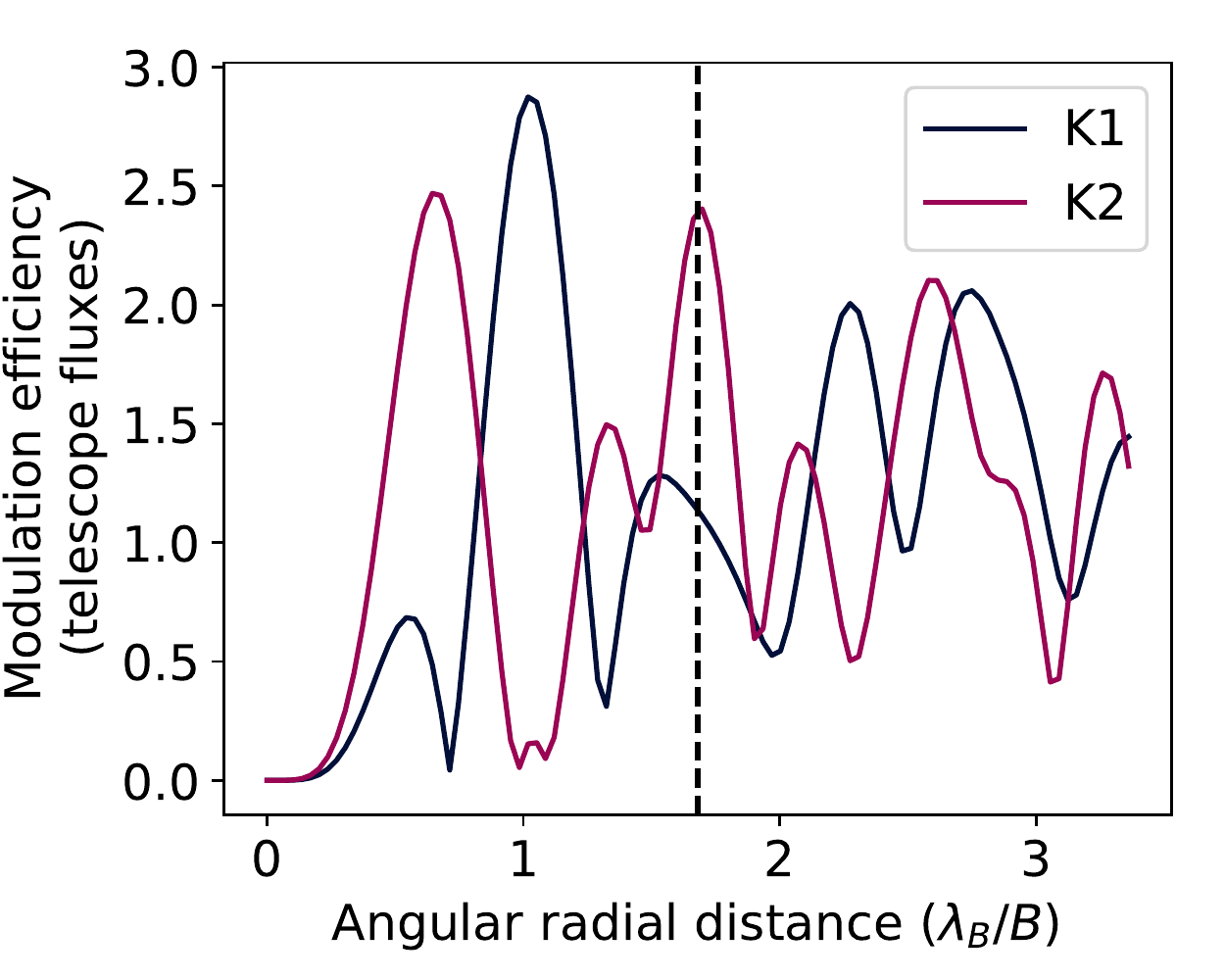}  \\
         c & \includegraphics[width = \linewidth]{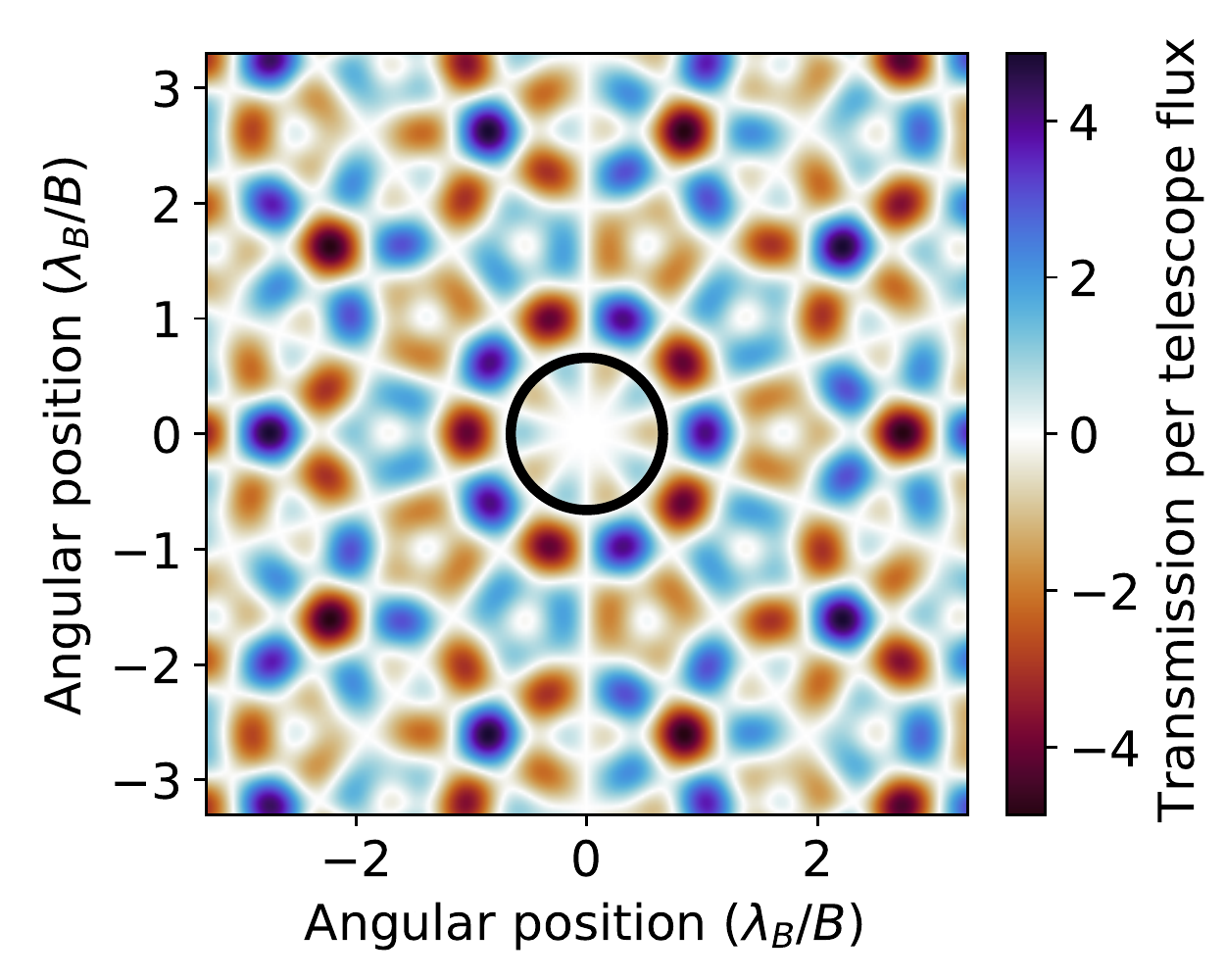}& \includegraphics[width = \linewidth]{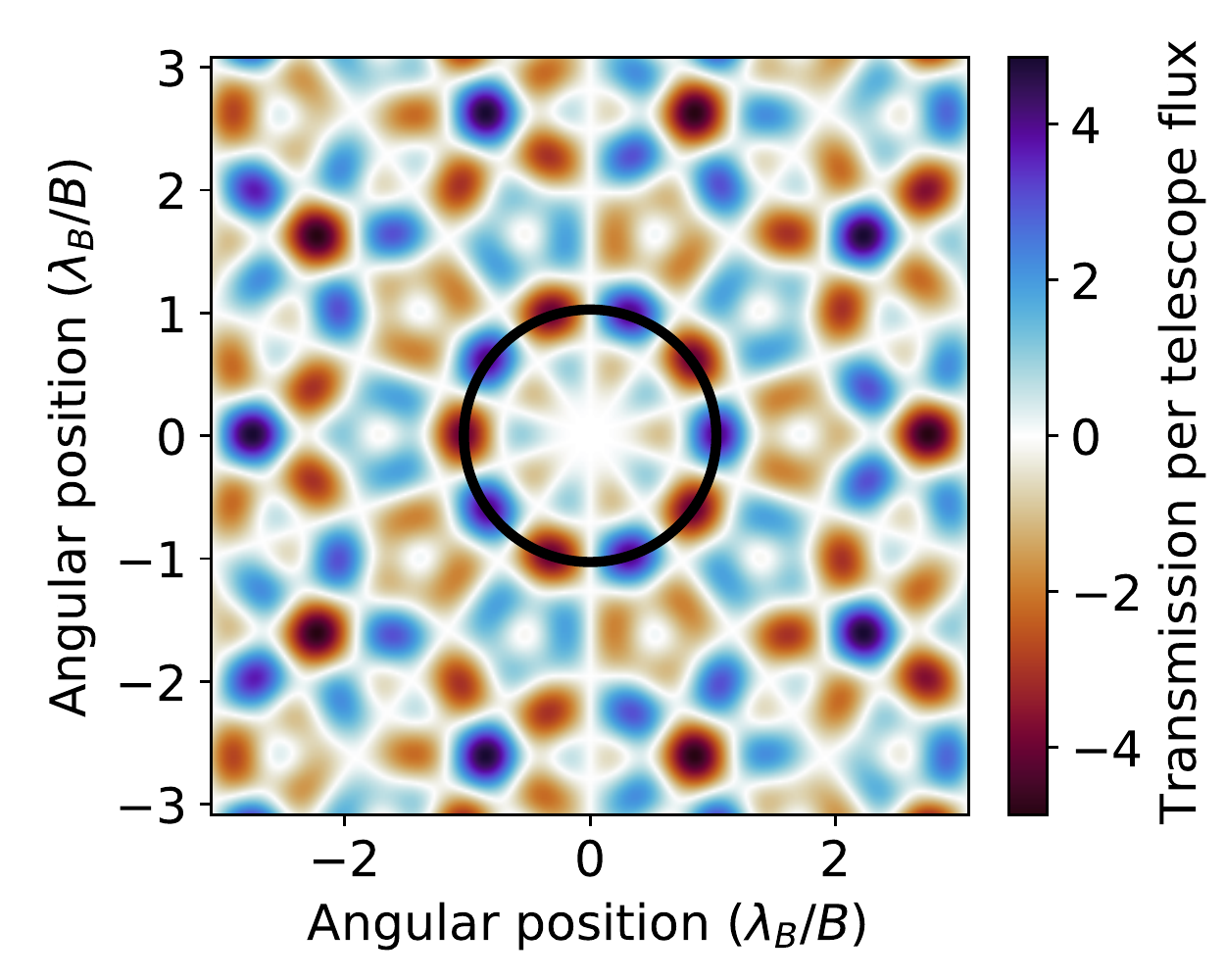} & \includegraphics[width = \linewidth]{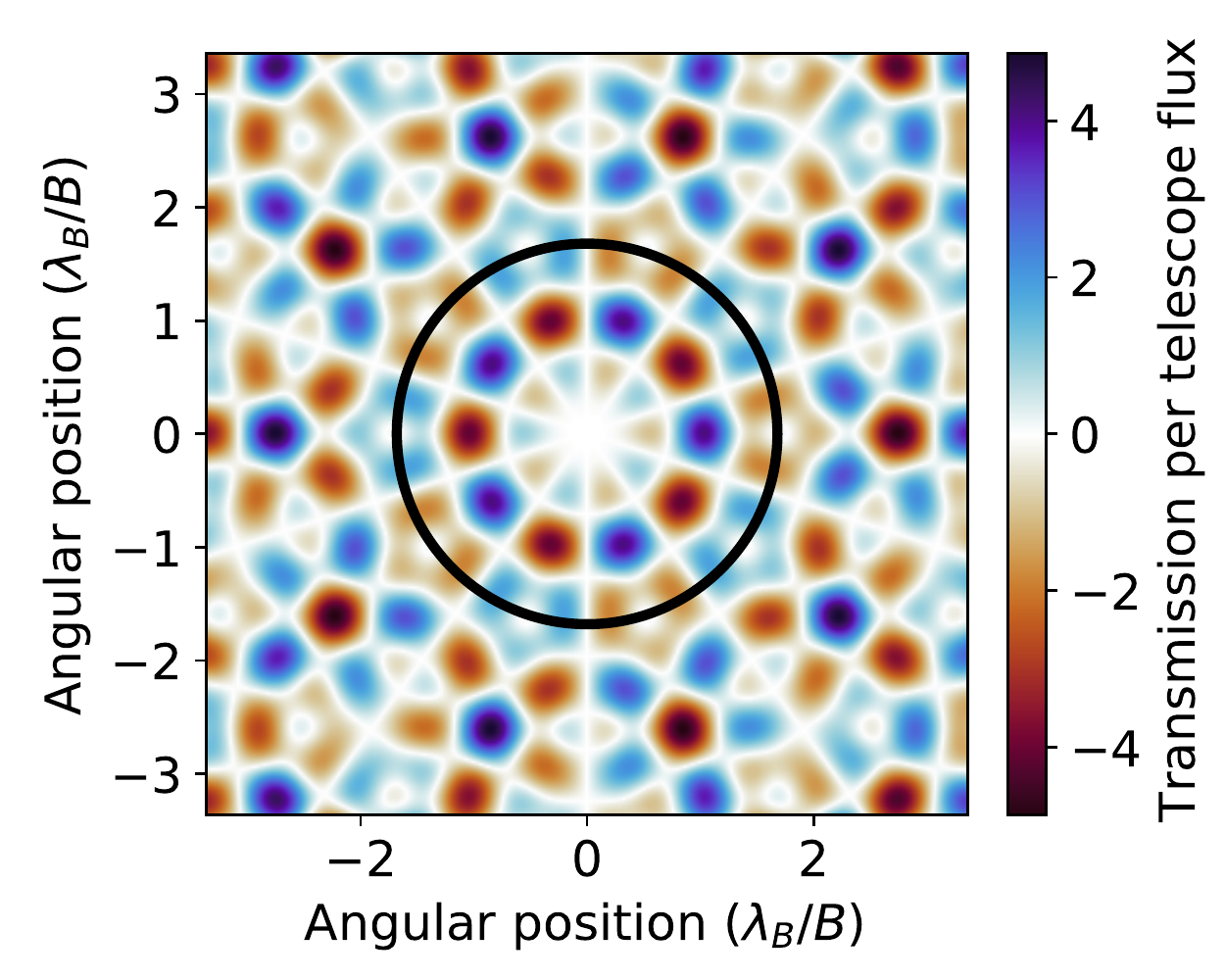} \\
         & \includegraphics[width = \linewidth]{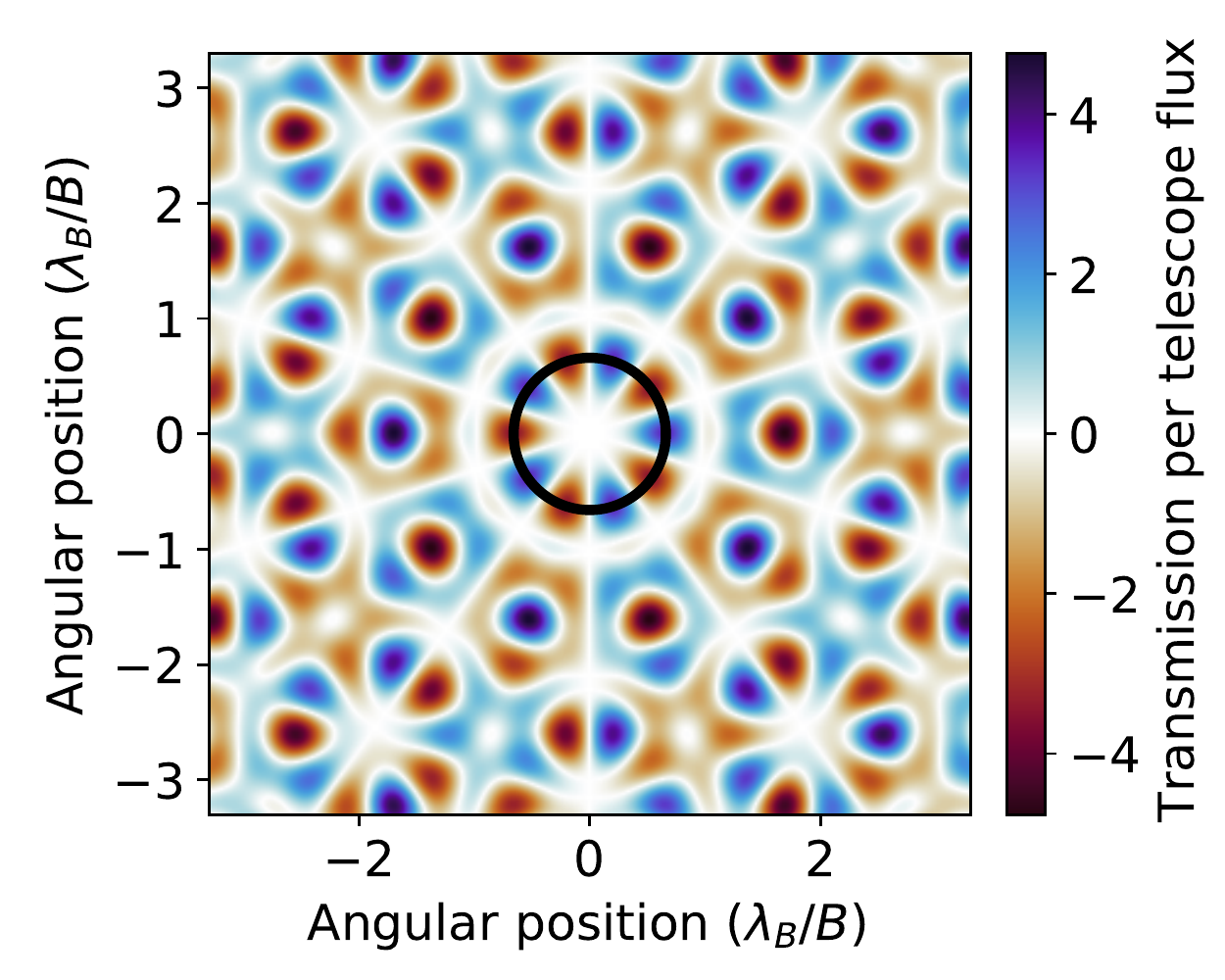}& \includegraphics[width = \linewidth]{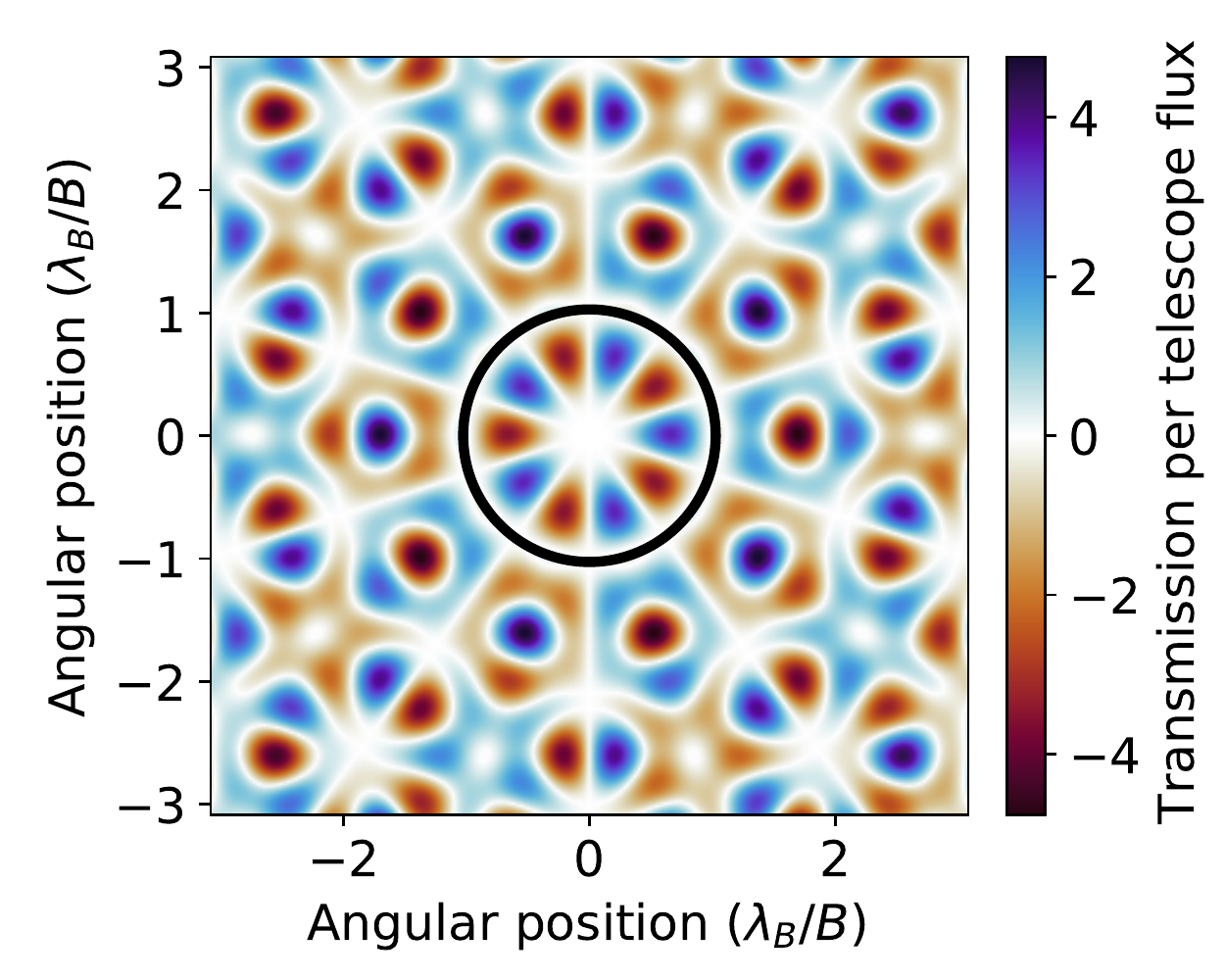} & \includegraphics[width = \linewidth]{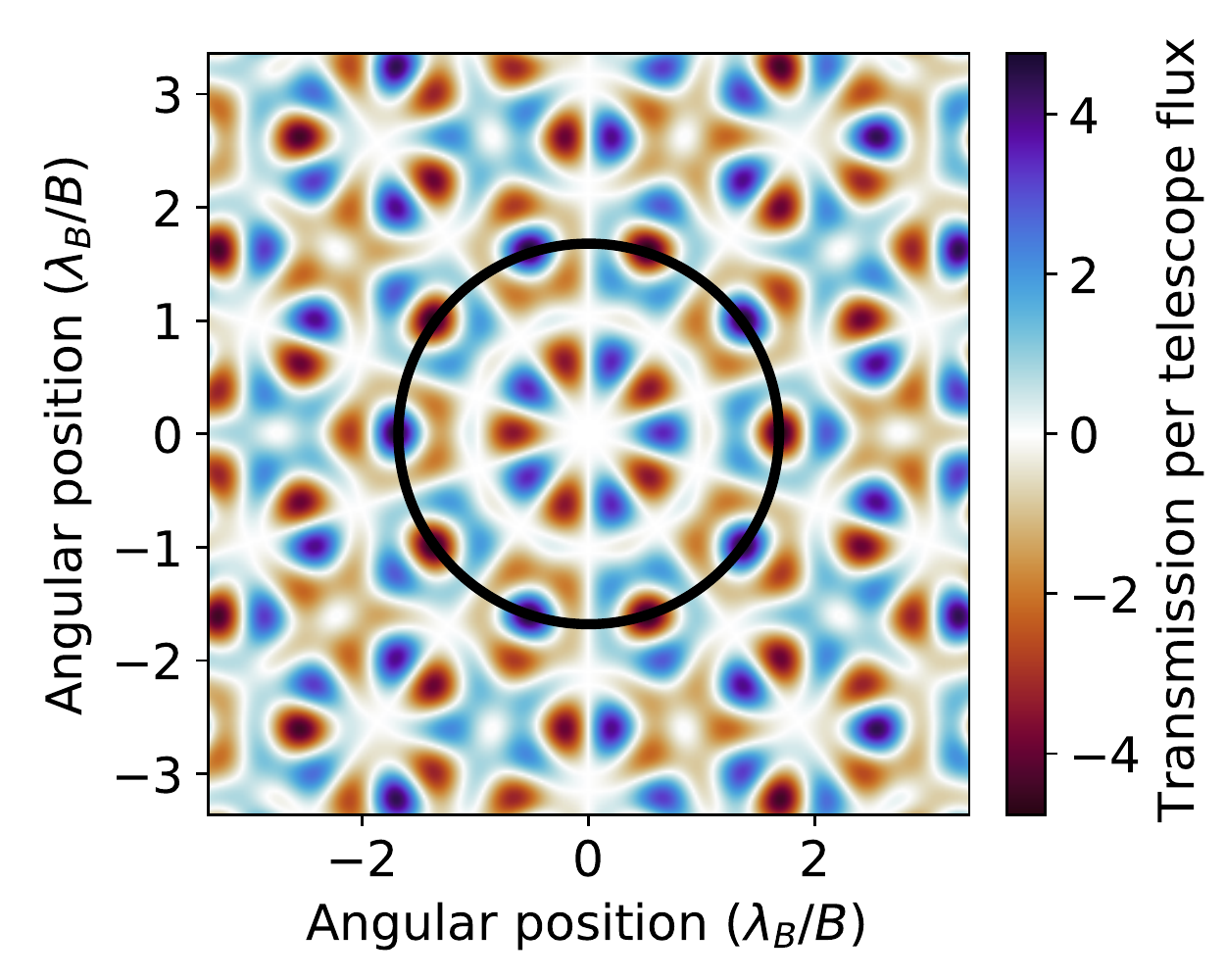} \\ \\
        \bottomrule
    \end{tabular}
    \caption{\red{Telescope configurations (subfigures a), modulation efficiency curves as a function of radial position for each kernel (subfigures b), and kernel maps (subfigures c) for each of the simulated architectures (X-array (A), Kernel-3 (B), Kernel-4 (C), and three different scalings of the Kernel-5 (D-F)). The dashed line on the modulation efficiency plots, and the circle on the kernel maps, correspond to the angular separation that the array is optimised for, and defines the value of $\Gamma_B$. In general, this corresponds to the angular separation with the highest modulation efficiency at the reference wavelength. Angular position in these plots is given in units of $\lambda_B/B$, and the transmission is given in units of single telescope flux.}}
    \label{Architectures}
\end{figure*}

\subsubsection{Kernel-3 Nuller}

The Kernel-3 nuller arrangement is simply three telescopes in an equilateral triangle. This is similar to what was studied in the \textit{Darwin} era \citep{2004Karlsson}, although the beam combination scheme is different and is instead based on a kernel-nuller (see Sect. 5.1 in \cite{2020Laugier}). 

We can consider the Cartesian coordinates of the telescopes to be
\begin{align}
    x_k &\approx \frac{0.577 B\cos{\theta_k}}{\lambda_B} &
    y_k &\approx \frac{0.577 B\sin{\theta_k}}{\lambda_B},
\end{align}
where $\theta = [0, \frac{2\pi}{3}, \frac{4\pi}{3}]$ and $B$ is the baseline of adjacent spacecraft. A schematic of the design is shown in \red{Fig. \ref{Architectures}.B.a.}

The nulling transfer matrix is:
\begin{equation}
   \vb{M}_{k,l} = \frac{1}{\sqrt{3}}\begin{bmatrix}
    1 & 1 & 1\\
    1 & e^{\frac{2\pi i}{3}} & e^{\frac{4\pi i}{3}}\\
    1 & e^{\frac{4\pi i}{3}} & e^{\frac{2\pi i}{3}}.
\end{bmatrix}
\end{equation}
We note that there is a single kernel-null here ($N_K = 1$), given by the difference in the last two outputs, with the starlight going into the first output. We first normalise the transmission to flux per telescope,
\begin{align}
    \tilde{\vb{R}}_l(\vb*{\alpha}) = 3\frac{\vb{R}_l(\vb*{\alpha})}{\red{\vb{R}_0(0)}},
\end{align}
and the kernel-null response (and associated transmission map) is given by
\begin{align}
    \vb{K}_1(\vb*{\alpha}) &=  \tilde{\vb{R}}_1(\vb*{\alpha}) - \tilde{\vb{R}}_2(\vb*{\alpha}) & \vb{T}_1(\vb*{\alpha}) = \red{\frac{1}{2}\left(\tilde{\vb{R}}_1(\vb*{\alpha})+\tilde{\vb{R}}_2(\vb*{\alpha})\right)}.
\end{align}

As with the X-array design, through calculating the \red{modulation efficiency for different radii plotted in Fig. \ref{Architectures}.B.b,} we identified that the baseline scale factor for this configuration should be $\Gamma_B = 0.666$. A map of the kernel-null response is shown in Fig. \red{\ref{Architectures}.B.c}.

\subsubsection{Kernel-4 Nuller}
The Kernel-4 nuller is the most interesting of the architectures, as it directly competes with the X-array design in terms of number of telescopes. \red{Instead of having only two nulled outputs, however, this design produces three kernel-null observables ($N_K = 3$) through the use of an extra optical mixing stage in the beam combination process \citep{2018Martinache}, potentially providing more information for signal demodulation. Furthermore, this design only has one bright output and so should make more use of the planet flux. Conversely, as the signal is split between multiple outputs, there will be less signal per output than the X-array.}

\red{This design requires non-redundant baselines}, and so rectangular geometries will not be appropriate. Instead, a right angled kite design was chosen due to it's ability to be parameterised around a circle. The kite is defined by two parameters as with the X-array design - the short baseline ($B$) and the ratio of the longer side to the shorter side ($c$). A schematic is in \red{Fig. \ref{Architectures}.C.a}. 

The azimuthal angular positions of the telescopes can be derived through the geometry of the right kite, yielding
\begin{align}
    \theta = \left[\frac{\pi}{2}-\beta, \frac{\pi}{2}, \frac{\pi}{2}+\beta,\frac{3\pi}{2}\right], &&
    \beta = 2\tan^{-1}\left(\frac{1}{c}\right),
\end{align}
and Cartesian coordinates of
\begin{align}
    x_k &= \frac{0.5B\sqrt{1 + c^2}\cos{\theta_k}}{\lambda_B} &
    y_k &= \frac{0.5B\sqrt{1 + c^2}\sin{\theta_k}}{\lambda_B}.
\end{align}

The beam combination transfer matrix, as given in \cite{2018Martinache}, is
\begin{equation}
   \vb{M}_{k,l} = \frac{1}{4}\begin{bmatrix}
    2 & 2 & 2 & 2\\
    1 + e^{i\theta} & 1 - e^{i\theta} & -1 + e^{i\theta} & -1 - e^{i\theta}\\
    1 - e^{-i\theta} & -1 - e^{-i\theta} & 1 + e^{-i\theta} & -1 + e^{-i\theta}\\
    1 + e^{i\theta} & 1 - e^{i\theta} & -1 - e^{i\theta} & -1 + e^{i\theta}\\
    1 - e^{-i\theta} & -1 - e^{-i\theta} & -1 + e^{-i\theta} & 1 + e^{-i\theta}\\
    1 + e^{i\theta} & -1 - e^{i\theta} & 1 - e^{i\theta} & -1 + e^{i\theta}\\
    1 - e^{-i\theta} & -1 + e^{-i\theta} & -1 - e^{-i\theta} & 1 + e^{-i\theta}\\
\end{bmatrix},
\end{equation}
where $\theta$ is the optical mixing angle, set to $\frac{\pi}{2}$. We note that there are four inputs and seven outputs in this design, accounting for the one bright output and three pairs of nulled outputs. Again, we normalise the output to be in terms of telescope flux:
\begin{align}
    \tilde{\vb{R}}_l(\vb*{\alpha}) = 4\frac{\vb{R}_l(\vb*{\alpha})}{\red{\vb{R}_0(0)}}.
\end{align}
The kernel-null responses are the difference between the adjacent null responses, and are listed with their associated transmission maps below:
\begin{align}
    \vb{K}_1(\vb*{\alpha}) &=  \tilde{\vb{R}}_1(\vb*{\alpha}) - \tilde{\vb{R}}_2(\vb*{\alpha}) & \vb{T}_1(\vb*{\alpha}) &= \red{\frac{1}{2}\left(\tilde{\vb{R}}_1(\vb*{\alpha})+\tilde{\vb{R}}_2(\vb*{\alpha})\right)}\\    \vb{K}_2(\vb*{\alpha}) &=  \tilde{\vb{R}}_3(\vb*{\alpha}) - \tilde{\vb{R}}_4(\vb*{\alpha}) & \vb{T}_2(\vb*{\alpha}) &= \red{\frac{1}{2}\left(\tilde{\vb{R}}_3(\vb*{\alpha})+\tilde{\vb{R}}_4(\vb*{\alpha})\right)}\\
    \vb{K}_3(\vb*{\alpha}) &=  \tilde{\vb{R}}_5(\vb*{\alpha}) - \tilde{\vb{R}}_6(\vb*{\alpha}) & \vb{T}_3(\vb*{\alpha}) &= \red{\frac{1}{2}\left(\tilde{\vb{R}}_5(\vb*{\alpha})+\tilde{\vb{R}}_6(\vb*{\alpha})\right)}.
\end{align}

Finally, as with the previous architectures, we identified the baseline scale factor through an analysis of the \red{modulation efficiency} as a function of radius (\red{Fig. \ref{Architectures}.C.B}). In particular, we found that the baseline scale factor is $\Gamma_B=0.4$ when the kite has a ratio of $c=1.69$. This ratio was chosen as the three kernel-null outputs have maxima in roughly the same place. In fact, due to the symmetry of the kite, kernels 1 and 3 are anti-symmetrical and have the exact same radial RMS response. This can be seen in the plots of the kernel-nulls in \red{Fig. \ref{Architectures}.C.c}.

\subsubsection{Kernel-5 Nuller}

The final architecture we consider is the Kernel-5 nuller - five telescopes arranged in a regular pentagonal configuration, shown in \red{Figs. \ref{Architectures}.D.a to \ref{Architectures}.F.a}. The positions of the telescopes are given by
\begin{align}
    x_k &\approx \frac{0.851 B\cos{\theta_k}}{\lambda_B} &
    y_k &\approx \frac{0.851 B\sin{\theta_k}}{\lambda_B},
\end{align}
where again $B$ is the separation between adjacent spacecraft/the short baseline, and $\theta = [0, \frac{2\pi}{5}, \frac{4\pi}{5},\frac{6\pi}{5}, \frac{8\pi}{5}]$.

The transfer matrix for this beam combiner is extrapolated from that of the Kernel-3 nuller:
\begin{equation}
   \vb{M}_{k,l} = \frac{1}{\sqrt{5}}\begin{bmatrix}
    1 & 1 & 1 & 1 & 1\\
    1 & e^{\frac{2\pi i}{5}} & e^{\frac{4\pi i}{5}} & e^{\frac{6\pi i}{5}} & e^{\frac{8\pi i}{5}}\\
    1 & e^{\frac{4\pi i}{5}} & e^{\frac{8\pi i}{5}} & e^{\frac{2\pi i}{5}} & e^{\frac{6\pi i}{5}}\\
    1 & e^{\frac{6\pi i}{5}} & e^{\frac{2\pi i}{5}} & e^{\frac{8\pi i}{5}} & e^{\frac{4\pi i}{5}}\\
    1 & e^{\frac{8\pi i}{5}} & e^{\frac{6\pi i}{5}} & e^{\frac{4\pi i}{5}} & e^{\frac{2\pi i}{5}}.
\end{bmatrix}
\end{equation}
Here, we have one bright output and two pairs of nulled outputs that can produce kernel-nulls ($N_K = 2$). Again, we normalise the outputs to that of one telescope flux:
\begin{align}
    \tilde{\vb{R}}_l(\vb*{\alpha}) = 5\frac{\vb{R}_l(\vb*{\alpha})}{\red{\vb{R}_0(0)}}.
\end{align}
The two kernel-nulls and their associated transfer maps are given by
\begin{align}
    \vb{K}_1(\vb*{\alpha}) &=  \tilde{\vb{R}}_1(\vb*{\alpha}) - \tilde{\vb{R}}_4(\vb*{\alpha}) & \vb{T}_1(\vb*{\alpha}) &= \red{\frac{1}{2}\left(\tilde{\vb{R}}_1(\vb*{\alpha})+\tilde{\vb{R}}_4(\vb*{\alpha})\right)}\\
    \vb{K}_2(\vb*{\alpha}) &=  \tilde{\vb{R}}_2(\vb*{\alpha}) - \tilde{\vb{R}}_3(\vb*{\alpha}) & \vb{T}_2(\vb*{\alpha}) &= \red{\frac{1}{2}\left(\tilde{\vb{R}}_2(\vb*{\alpha})+\tilde{\vb{R}}_3(\vb*{\alpha})\right)}.
\end{align}

Now, after the same \red{modulation efficiency} analysis, it was found that there were maximum peaks at different places for each of the two kernel-nulls. Due to this, we simulated three different versions of the Kernel-5 nuller with different values for the baseline scaling. These values are $\Gamma_B = 0.66, 1.03$ and 1.68, and are shown \red{overlaid on the modulation efficiency curves in Figs. \ref{Architectures}.D.b to \ref{Architectures}.F.b respectively}. In the forthcoming analysis, these arrangements will be distinguished through the value of their baseline scale factor.  \red{The kernel maps of each of these scaled versions of the Kernel-5 array are found in Figs \ref{Architectures}.D.c to \ref{Architectures}.F.c}.

\subsection{Signal and Noise sources}

To determine whether a planet is detectable, or to determine the extent by which a planet can be characterised in a certain amount of time, we use the signal to noise ratio (SNR) metric. In our analysis, we are assuming we are photon noise limited and hence only consider photon noise sources. Considering fringe tracking noise would require a model of the power spectrum noise in the servo loop, which is highly dependent on the architecture and beyond the scope of this paper, although it will be briefly considered in the discussion. The sources we include are stellar leakage, the starlight that seeps past the central null; zodiacal light, thermal emission of dust particles in our solar system; and exozodiacal light, the equivalent of zodiacal light but from the target planet's host star. Each $i$th spectral channel and $j$th kernel-null can be calculated using
\begin{equation}
    SNR_{i,j} = \sqrt{\frac{\eta t}{2}}\frac{F_{\text{signal},i,j}A}{\sqrt{F_{\text{leakage},i,j}A + F_{\text{exozod},i,j}A + \red{a}P_{\text{zod},i}}},
\end{equation}
where $\eta$ is the throughput of the interferometer, $t$ is the exposure time, $A$ is the telescope area, $F_{\text{signal},i,j}$ is the planet flux, $F_{\text{leakage},i,j}$ is the stellar leakage flux, $F_{\text{exozod},i,j}$ is the exozodiacal light flux and $P_{\text{zod},i}$ is the zodiacal power. Each of these flux values are further defined below in the following sections. \red{The parameter $a$ is used as a scaling factor for the zodiacal light of the Kernel-4 kite configuration; due to the extra split, the zodiacal light is reduced by a factor of two. Hence $a=0.5$ for the Kernel-4, and $a=1$ for the other configurations.} The zodiacal light is a background source and is thus not dependant on the telescope area. The factor of $\frac{1}{\sqrt{2}}$ included in the calculation stems from the use of the difference of two nulled maps for the signal. We then combine the separate spectral channel SNRs and different kernel map SNRs through \cite{2004LayNoise}:
\begin{equation}
    SNR_{tot} = \sqrt{\sum_{i,j} SNR_{i,j}^2}.
\end{equation}

\red{Before addressing the separate noise sources used in the simulation, we mention briefly about the monochromatic sensitivity of these configurations at a known star-planet separation and position angle. In the case of the X-array, Kernel-3 and Kernel-5, there are spatial locations where the kernel map has a transmission amplitude equal to the number of telescopes. This means that in the case of known star-planet separation, for a single wavelength, all the planet light comes out one output, and the background is split between all outputs. In turn this means that for the same total collecting area, they have identical SNR in the case of zodiacal and photon noise. This is a maximum signal to noise in the case of a background-limited chopped measurement. Hence the main differences between these architectures comes from their relative polychromatic responses (and modulation efficiencies in the case of array rotation and planet detection), and their sensitivity to different noise sources, notably the depth of the null with regards to stellar leakage. For the Kernel-4, due to the transmission maxima of one output never occurring at the minima of other outputs, the maximum SNR is 8\% lower than the X-array's theoretical maximum with four telescopes. As an aside, if no nulling was required and a single telescope of the same collecting area was used in an angular chopping mode, it would also have the same signal to noise (albeit with a slightly different architecture, as two neighbouring sky positions would be simultaneously recorded while chopping).}

We now explain our model for the signal and each of these noise sources. 

\subsubsection{Planet signal}
\label{Sec:planet_signal}
The signal of the planet depends on which mode of observation we are undergoing - search or characterisation.

\red{Recall that the search mode is where the array is optimised for a single point around a star, and made to rotate so that any planet signal is modulated. } We have chosen the angle at which the array is optimised as the centre of the habitable zone (HZ). We calculate the HZ distances using the parameterisation of \cite{2013Kopparapu}, with the coefficients for the outer edge being `Early Mars' and the inner edge being `Recent Venus'. Specifically, we have that the stellar flux at the inner edge and outer edge of the HZ are given by
\begin{align}
    S_\text{in} &= S_{0,\text{in}} + A_\text{in}T + B_\text{in}T^2 + C_\text{in}T^3\\
    S_\text{out} &= S_{0,\text{out}} + A_\text{out}T + B_\text{out}T^2 + C_\text{out}T^3,
\end{align}
where $T$ is the effective temperature of the star minus the Sun's temperature ($T = T_\text{eff} - 5780$\,K) and
\begin{align*}
    S_{0,\text{in}} &= 1.7665 & S_{0,\text{out}} &= 0.3240\\
    A_\text{in} &= 1.3341\times 10^{-4} & A_\text{out} &= 5.3221\times 10^{-5}\\
    B_\text{in} &= 3.1515\times 10^{-9} & B_\text{out} &= 1.4288\times 10^{-9}\\
    C_\text{in} &= -3.3488\times 10^{-12} & C_\text{out} &= -1.1049\times 10^{-12}.
\end{align*}
We scale the stellar flux using the bolometric luminosity of the star, $L_\text{Bol}$, to obtain the radius:
\begin{align}
    r_\text{in} &= \sqrt{\frac{L_\text{Bol}}{S_{in}}}& r_{out} &= \sqrt{\frac{L_\text{Bol}}{S_{out}}}.
\end{align}

Finally, we adopt the mean of this range and divide by the stellar distance to obtain the angular position of the centre of the HZ. This position is used to set the baseline of the array, which in turn creates the response maps (as in Sect. \ref{Sec:architectures}). We then take the angular separation of the planet, $r_i$, scaled by the central wavelength of each channel, $\lambda_i$, and average the kernel-null observable map over azimuthal angles \red{(akin to the modulation efficiency)}. Specifically, the planet transmission in the kernel-nulled transmission map $\vb{K}_j$, $j\in[1,..,N_k]$ averaged over angles $\phi$ is
\begin{align}
    \red{T_{i,j} = \xi_j(r_i) = \sqrt{\langle \vb{K}_j(r_i,\phi)^2\rangle_\phi}}.
\end{align}
The planet signal is then just the transmission multiplied by the planet flux (consisting of both thermal and reflected contributions) in the different $i\in[1,..,N_\lambda]$ spectral channels:
\begin{align}
    F_{signal,i,j} = T_{i,j}F_{planet,i}.
\end{align}

For the characterisation mode, we assume we know precisely where the planet we are investigating is. Hence, the optimised angular separation is the same as the angular separation of the planet at the characterisation wavelength ($\delta = r_{\lambda_B}$). To identify the best azimuthal angle of the array to ensure the highest signal, we find the azimuthal angle that gives the maximum summed transmission over wavelengths:
\begin{align}
    \red{\phi_{j,\text{max}} = \max_{\{\phi\}}\sum_i^{N_\lambda}{\vb{K}_j(r_i,\phi)}}.
\end{align}
This angle is then used to calculate the \red{wavelength dependent} transmission of the planet, which when multiplied by the planet flux gives the signal:
\begin{align}
    \red{F_{\text{signal},i,j} = \vb{K}_j\left(r_i,\phi_{j,\text{max}}\right)F_{\text{planet},i}}.
\end{align}

\subsubsection{Stellar Leakage}
Stellar leakage is the light from the star that `leaks' into the nulled outputs. The stellar leakage flux is the summed grid of the stellar flux (for each spectral channel $i$) multiplied by a normalised limb darkening law $I(r)$ and the $j$th transmission function:
\begin{align}
    F_{\text{leakage},i,j} = \sum_{m,n}\vb{T}_j(\vb*{\alpha}_{m,n,i})\frac{I\left(\vb{r}_{m,n,i}\right)}{\sum_{m,n} I\left(\vb{r}_{m,n,i}\right)}F_{\text{star},i},
\end{align}
where $\vb{r}_{m,n,i}$ is the linear coordinate in units of stellar radius \red{$\theta$} given by
\begin{equation}
  \vb{r}_{m,n,i} = \frac{\vb*{\alpha}_{m,n,i}}{\theta}.
\end{equation}

We use a standard second order limb darkening law where $r$ is the radius from the centre of the transmission map (in units of stellar radius). Coefficients are from \cite{2011Claret}, using the Spitzer 8~\textmu m filter and assuming a typical K dwarf with $T=5000$~K, log(g)$=4.5$ and [Fe/H] = 0:
\begin{align}
    I(r) = 1-0.05\left(1-\sqrt{1-r^2}\right)-0.10\left(1-\sqrt{1-r^2}\right)^2.
\end{align}

\subsubsection{Zodiacal Light}

The zodiacal light is the primary background source for space telescopes, being the light reflected (and thermally radiated) from dust in the solar system. To calculate this, we use the JWST background calculator\footnote{GitHub: \url{https://github.com/spacetelescope/jwst_background}}, as this telescope operates in the same wavelength range. 

To begin with, we find the zodiacal light spectral radiance in the direction of the star's celestial coordinates. The JWST calculator returns the spectral radiance expected over the course of a year - this is due to the different position of the sun with respect to the telescope. Due to the Emma array type design, with the beam combiner out of plane of the collector telescopes, we assume that the interferometer can only look in an anti-solar angle from 45 to 90 degrees. On average, this corresponds roughly to the 30\% percentile of the yearly distribution, and so we adopt this radiance.

We note that, assuming the background over the field of view is isotropic, the solid angle subtended by the telescope PSF is proportional to
\begin{align}
    \Omega \propto \frac{\lambda^2}{A},
\end{align}
where $A$ is the telescope aperture area. Hence we can convert the spectral radiance of the zodiacal light to spectral power ($ph/s/m$) by multiplying by the square of the wavelength:
\begin{align}
    P_\text{zod}(\lambda) = \lambda^2L_\text{zod}(\lambda).
\end{align}

We then calculate the background for each $i$th spectral channel by integrating over wavelength:
\begin{equation}
    P_{\text{zod},i} = \int_{\lambda_{i,\text{min}}}^{\lambda_{i,\text{max}}} P_\text{zod}(\lambda)d\lambda.
\end{equation}

\subsubsection{Exozodiacal Light}
Exozodiacal light is simply the equivalent to zodiacal light around interstellar systems. To calculate this, we will simply scale the local zodiacal background by a number of `exozodis', calculated through P-Pop from the distributions in \cite{2020Ertel}. One exozodi is equivalent to the zodiacal light in our own solar system. We assume the exozodiacal background is not clumpy, and is distributed face on (with an inclination of zero). These assumptions may not be realistic, but further analysis of the complications of exozodiacal light is beyond the scope of this work. We point the reader to \cite{2010Defrere} for a better treatment of this background source. 

We start by calculating the local zodiacal spectral radiance as seen looking at the ecliptic pole using the JWST calculator as before with the zodiacal background calculation. Unlike before, we take the minimum value over the course of the year. We then integrate over each spectral channel, giving us the radiance in each channel:
\begin{equation}
    L_{\text{zod},\text{min},i} = \int_{\lambda_{i,\text{min}}}^{\lambda_{i,\text{max}}} L_{\text{zod},\text{min}}(\lambda)d\lambda.
\end{equation}

Now, we convert our angular sky grid $\vb*{\alpha}_{m,n}$ into a linear grid with units of AU, again scaling by the channel's central wavelength:
\begin{equation}
    \vb{r}_{m,n,i} = d\vb*{\alpha}_{m,n}\frac{\lambda_i}{\lambda_B},
\end{equation}
where $d$ is the distance to the star in AU and $\vb*{\alpha}_{m,n}$ is the angular coordinate in radians. We overlay the radial surface density distribution of zodiacal dust, which is assumed to scale with heliocentric distance as a power law with exponent 0.34 \citep{2015Kennedy}. We also factor in a density depletion factor $\lambda(r)$, as proposed by \cite{2021Stenborg}, to account for the depletion of dust as the heliocentric distance decreases. That is,
\begin{align}
    N(r) \propto \lambda(r)r^{-0.3},
\end{align}
where
\begin{align}
    \lambda(r) = \begin{cases}
    0 \quad &\text{if} \, r < r_\text{in} \\
    \frac{r-r_\text{in}}{r_\text{out}-r_{in}} \quad &\text{if} \, r_\text{in} \leq r \leq r_\text{out} \\
    1 \quad &\text{if} \, r > r_\text{out} 
    \end{cases}
\end{align}
for $r_\text{in} = 3\,R_\odot, r_\text{out} = 19\,R_\odot$. 

We then calculate the flux distribution through the Planck function. The temperature distribution of the dust is scaled by
\begin{align}
    T(r) = 300\text{\,K}\left(\frac{r}{1AU}\right)^{-0.5},
\end{align}
assuming the dust at 1\,AU is at 300\,K. Assuming blackbody radiation, the dust flux scales as
\begin{align}
    B(r,\lambda) \propto \frac{2\pi c}{\lambda^4}\frac{1}{e^{\frac{hc}{\lambda k T(r)}}-1}.
\end{align}
The flux distribution then, normalised at a radial distance of 1\,AU for each spectral channel is
\begin{align}
    F_i(r) = \gamma(r)r^{-0.3}\frac{\int_{\lambda_{i,\text{min}}}^{\lambda_{i,\text{max}}}B(r,\lambda)d\lambda}{\int_{\lambda_{i,\text{min}}}^{\lambda_{i,\text{max}}}B(1,\lambda)d\lambda}.
\end{align}

To get the scaled flux distribution, we multiply by the local scale factor at 1\,AU, given by:
\begin{align}
    s_i = 2zL_{\text{zod},\text{min},i},
\end{align}
where $z$ is the number of exozodis (from P-Pop). The factor of two arises from looking through two halves of the exozodiacal disk (as opposed to only looking through half of the local disk). Finally, we calculate the exozodiacal flux in each channel by multiplying the dust distribution by the $j$th transmission map ($\vb{T}_j$), the solid angle of each pixel ($\Omega$) and the local scale factor, and then summing over the grid:
\begin{align}
    F_{\text{exozod},i,j} = \sum_{m,n} F_i(\vb{r}_{m,n,i})s_i\vb{T}_j(\vb*{\alpha}_{m,n})\Omega.
\end{align}

\section{Results and discussion}

\subsection{Search Phase}
\label{Sec:results_search}
We begin by showing the number of detected exoplanets in the habitable zone for each of the six architectures, and also over three different reference wavelengths ($\lambda_B = 10,15$ and 18~\textmu m). We define an exoplanet as being detected when the total SNR over all the wavelength channels is greater than seven; the same as defined in LIFE1. Due to the technical challenges and physical restrictions that are imposed on a space interferometer, we set a limit to the possible baselines available. First, no two spacecraft can form a baseline shorter than 5~m, else they run the risk of colliding into each other. Secondly, we set that the baselines cannot extend beyond 600~m; this is where formation flying metrology may become more difficult and is a number used in the initial LIFE estimates (LIFE1). For any configurations outside of these limits, we set the SNR to be zero.

Unlike in LIFE1, we do not optimise integration time for our target list; instead we derive our estimates assuming a 5~hr integration time for every target. This number was chosen as a balance between the average integration times used in previous studies (15-35~hrs in LIFE1, 10~hrs in \cite{2018Kammerer}) and avoiding detection saturation of the targets within 20~pc. We emphasise that due to the difference in integration times, combined with the different underlying planet population along with a different treatment of the zodiacal and exozodiacal light, direct comparison of our estimates with the estimates in LIFE1 is not straightforward and should be treated with care. The purpose of this study is not the raw count of exoplanets detected, but rather the relative performance between architectures.

\begin{figure}
    \centering
    \includegraphics[width = 0.9\linewidth]{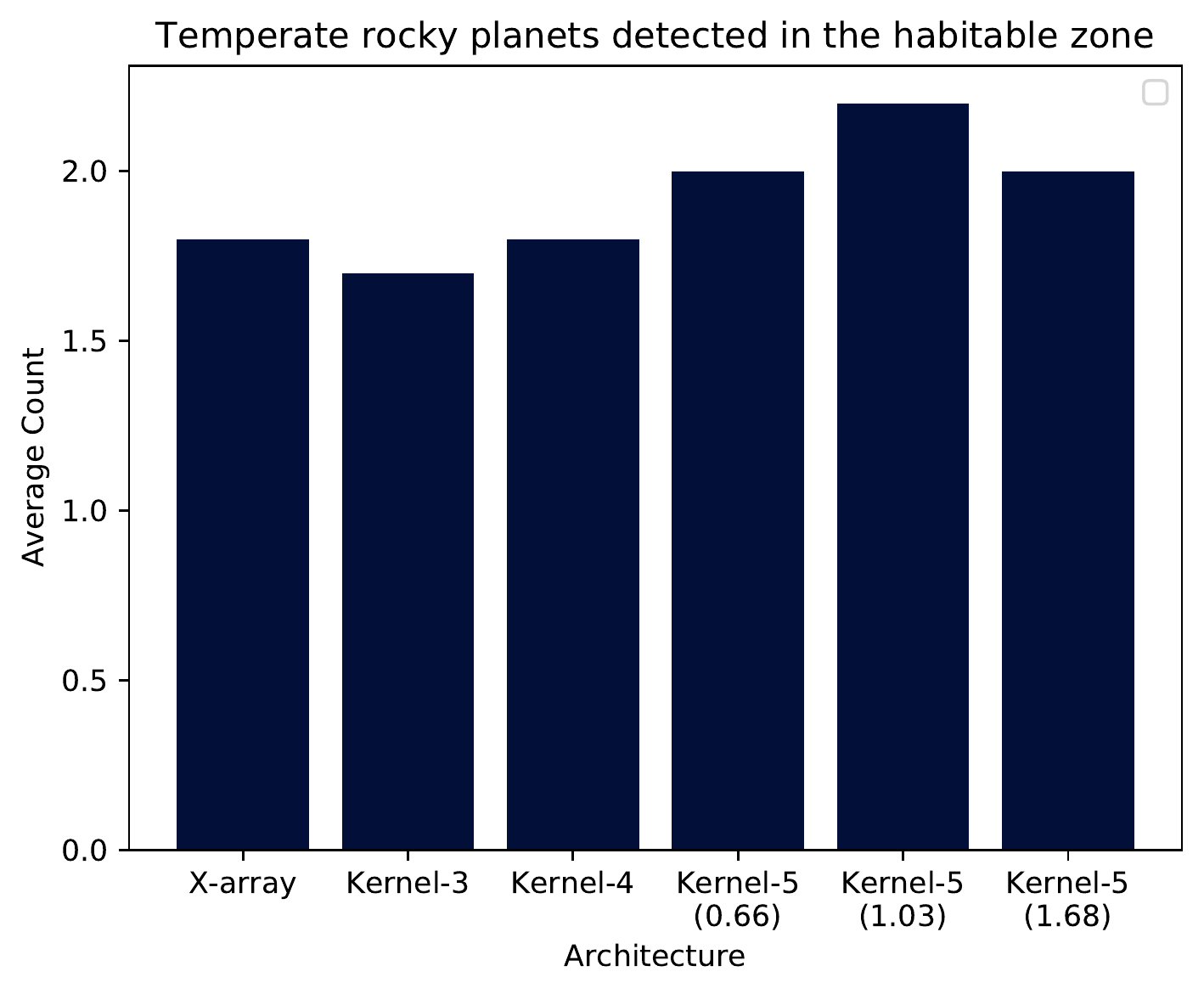}
    \caption{Number of Earth twins detected for each nulling architecture using a reference wavelength of 18~\textmu m. An Earth twin is defined as being rocky ($R < 1.5~R_\oplus$), temperate ($250 < T < 350$~K) and in the habitable zone.}
    \label{Img:Bar_habitable_rocky}
\end{figure}

We assume an optical throughput of the interferometer of 5\% and scale each collector's diameter such that the array as a whole has the same total area equal to four 2~m diameter telescopes. That is, the Kernel-3 nuller will have larger collectors, while the Kernel-5 nuller will have smaller collectors; this removes to first order the effect that architectures with more apertures will have a larger collecting surface. We divide the number of detections by the number of simulated universes to obtain the average count for one universe's worth of planets. The plots are shown in Fig. \ref{Img:Bar_habitable_rocky}.

\begin{figure}
  \centering
  \begin{subfigure}{0.6\linewidth} 
    \centering
    \includegraphics[width=\linewidth]{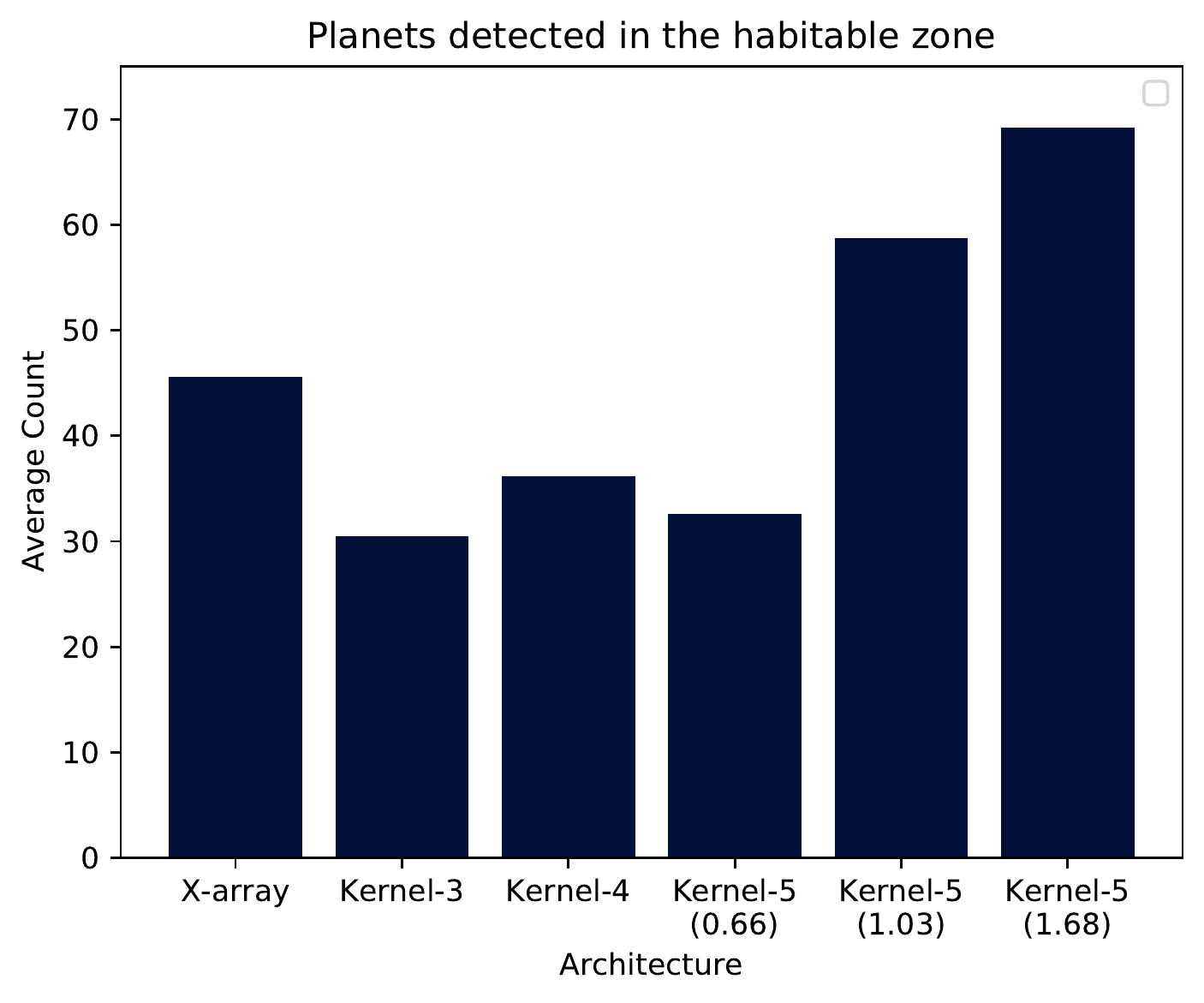}
    \caption{$\lambda_B = 10$~\textmu m}
  \end{subfigure}
  \hfill
  \begin{subfigure}{0.6\linewidth}
    \centering
    \includegraphics[width=\linewidth]{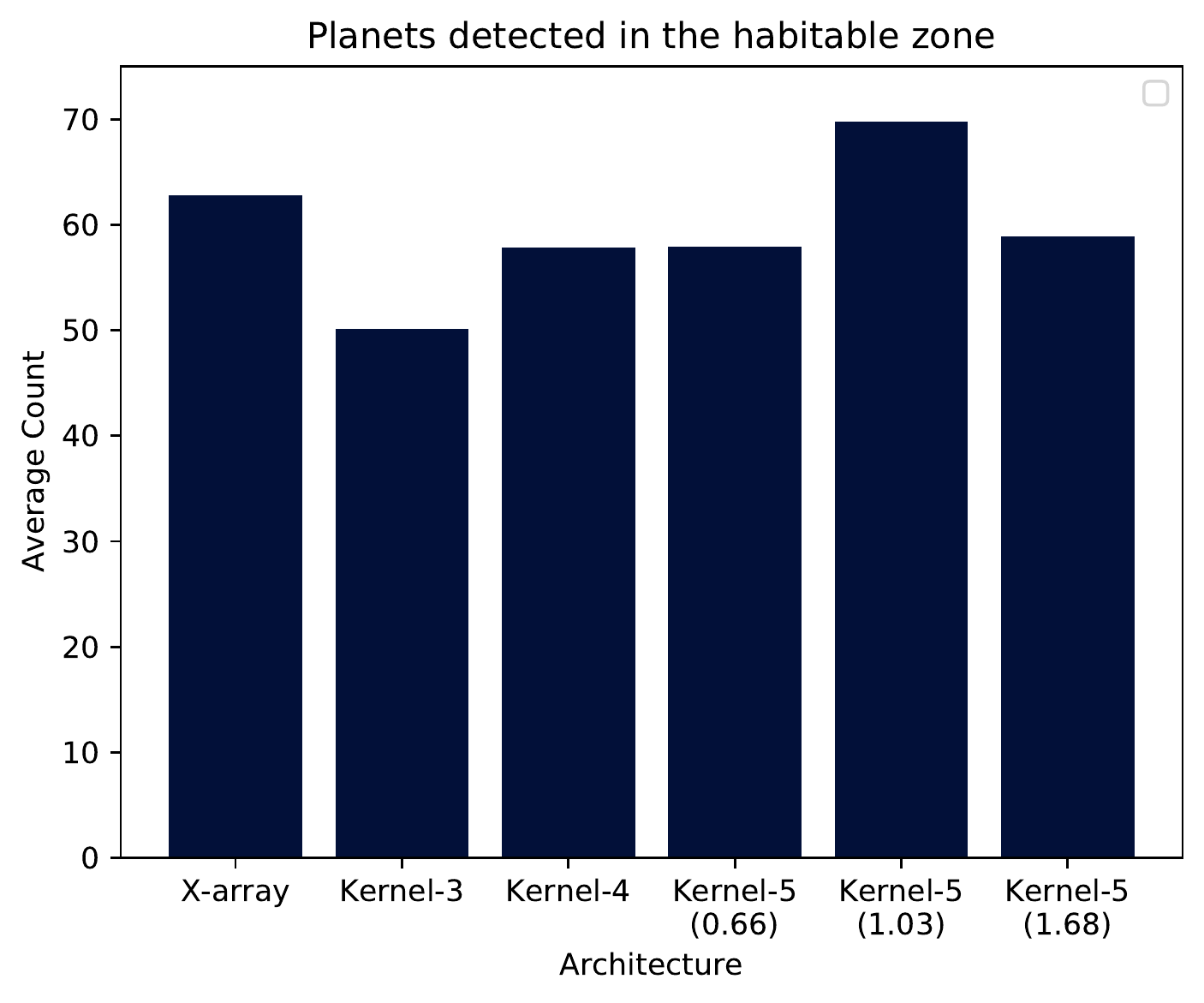}
    \caption{$\lambda_B = 15$~\textmu m}
  \end{subfigure}
    \begin{subfigure}{0.6\linewidth}
    \centering
    \includegraphics[width=\linewidth]{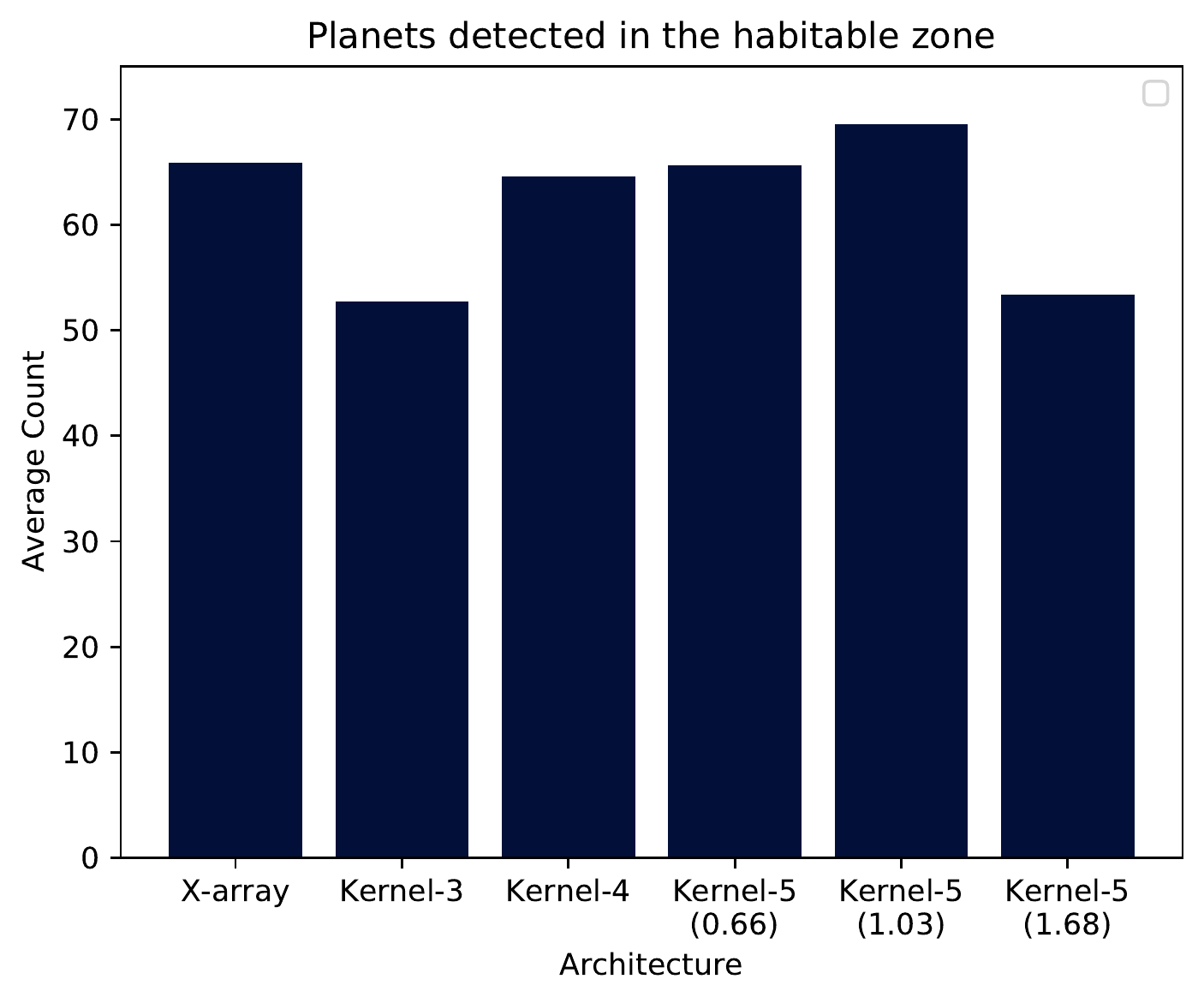}
    \caption{$\lambda_B = 18$~\textmu m}
  \end{subfigure}
    \caption{Number of planets detected in the habitable zone for each architecture, given as an average of the number of detections in each of the ten simulated universes. Each subfigure shows the detections for three different reference wavelengths ($\lambda_B$).}
    \label{Img:Bar_habitable}
\end{figure}

We can see that the Kernel-5 nuller, particularly the ones with a larger scale factor, perform the best in detecting habitable zone planets over all three reference wavelengths. The difference is most stark with $\lambda_B=10$~\textmu m, where the Kernel-5 nuller detects approximately 50\% more planets the X-array design. This advantage is then lowered at $\lambda_B=18$~\textmu m, where the X-array detects only a few less than the Kernel-5 (with $\Gamma_B = 1.03$). As generally the most detections over all the architectures occur when $\lambda_B=18$~\textmu m, we will restrict our analysis in the remaining sections to this reference wavelength. We also note here that our results echo that of \cite{LIFEPaper1} - reference wavelengths between 15 and 20~\textmu m produce similar yields for the X-array configuration, but substantially less at lower wavelengths.

\begin{figure}
    \centering
    \includegraphics[width = 0.9\linewidth]{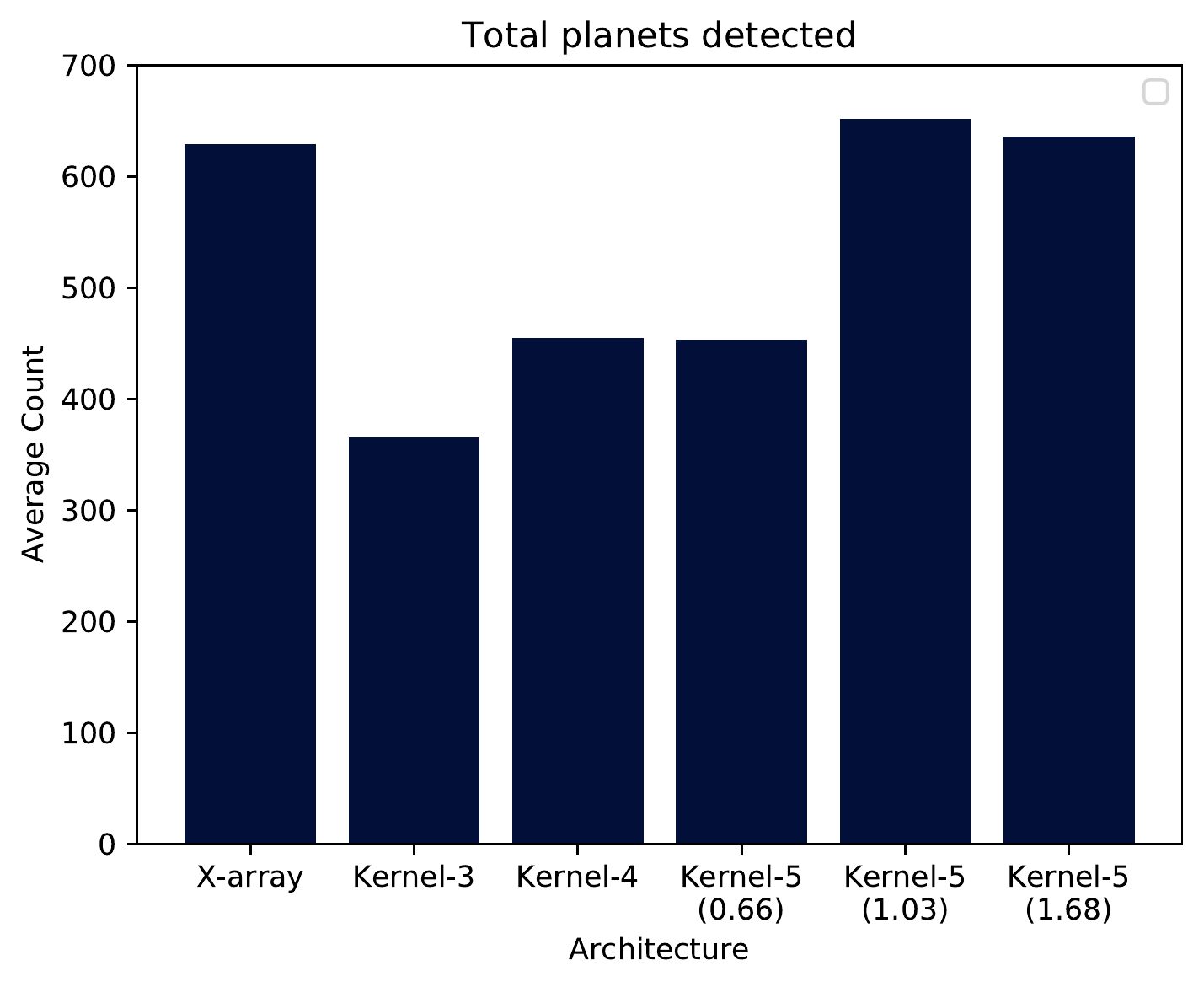}
    \caption{Total number of planets detected for each nulling architecture, for a reference wavelength of 18~\textmu m}
    \label{Img:Bar_total}
\end{figure}

We show in Fig. \ref{Img:Bar_total} the total planet yield of each of the architectures. We see again that the Kernel-5 architecture, particularly with a larger scaled baseline, detects substantially more planets than most of the architectures, although the X-array design provides a comparable yield. 

\begin{figure}
    \centering
    \includegraphics[width = 0.9\linewidth]{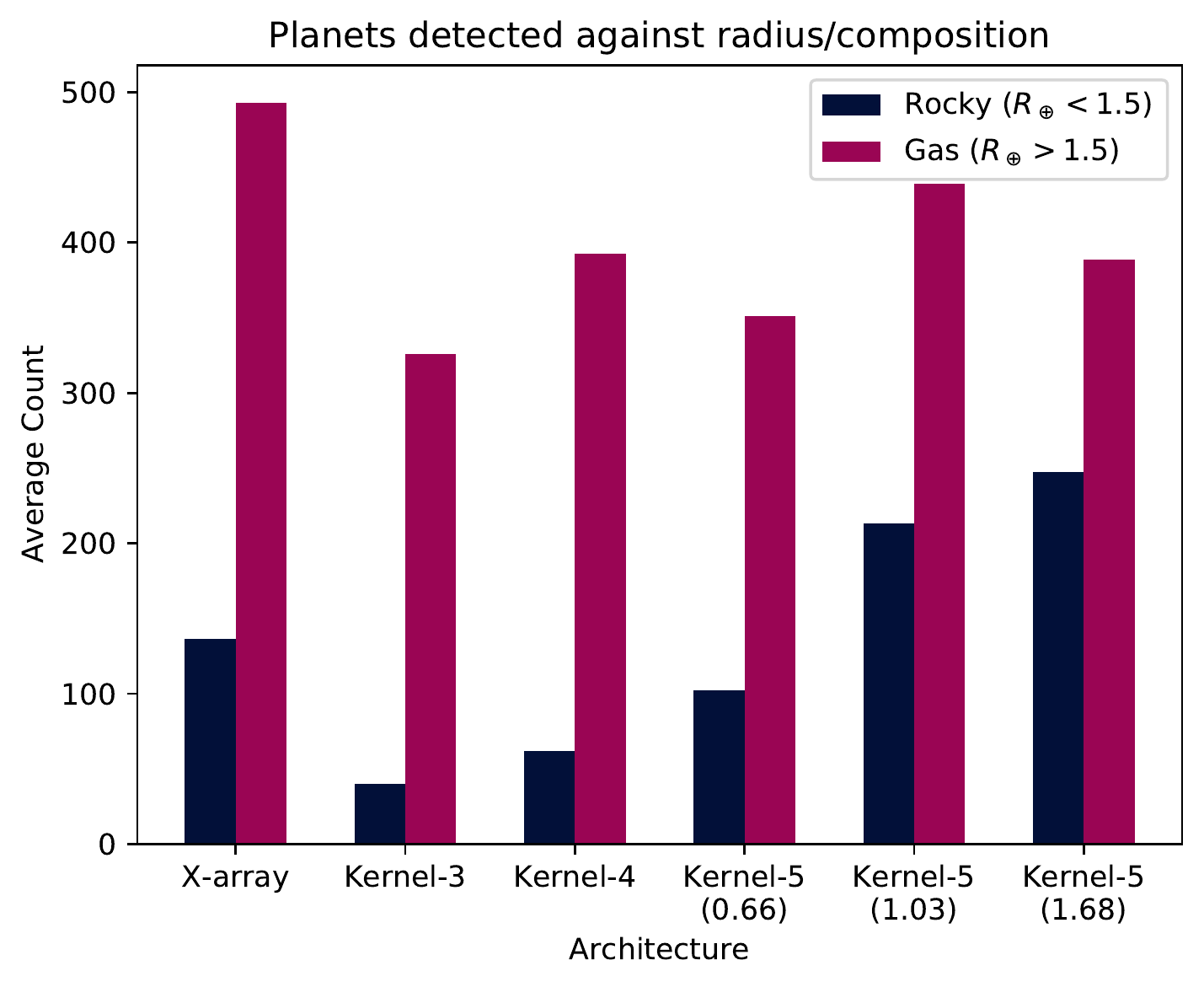}
      \caption{Number of planets detected for each nulling architecture, subdivided between two radii bins centred around 1.5~$R_\oplus$, for a reference wavelength of 18~\textmu m. This also acts as a proxy between rocky and gaseous planets.}
    \label{Img:Bar_radius}
\end{figure}

We also split up the exoplanet sample into a few categories. In Fig. \ref{Img:Bar_radius}, we split the detected sample into radii larger than and smaller than 1.5~$R_\oplus$. This value is used due to it being the point which is thought to separate rocky, super-Earths from gaseous sub-Neptunes \citep{2015Rogers,2017Chen}. Hence, this can be used as a rough metric of the composition of the planet. Interestingly, we see that the X-array detects mostly gaseous planets - more so than any of the other architectures, while the Kernel-5 nuller detects far more rocky planets; 60\% more than the X-array at $\Gamma_B = 1.03$ and 80\% more at $\Gamma_B = 1.68$. 

\begin{figure}
    \centering
    \includegraphics[width = 0.9\linewidth]{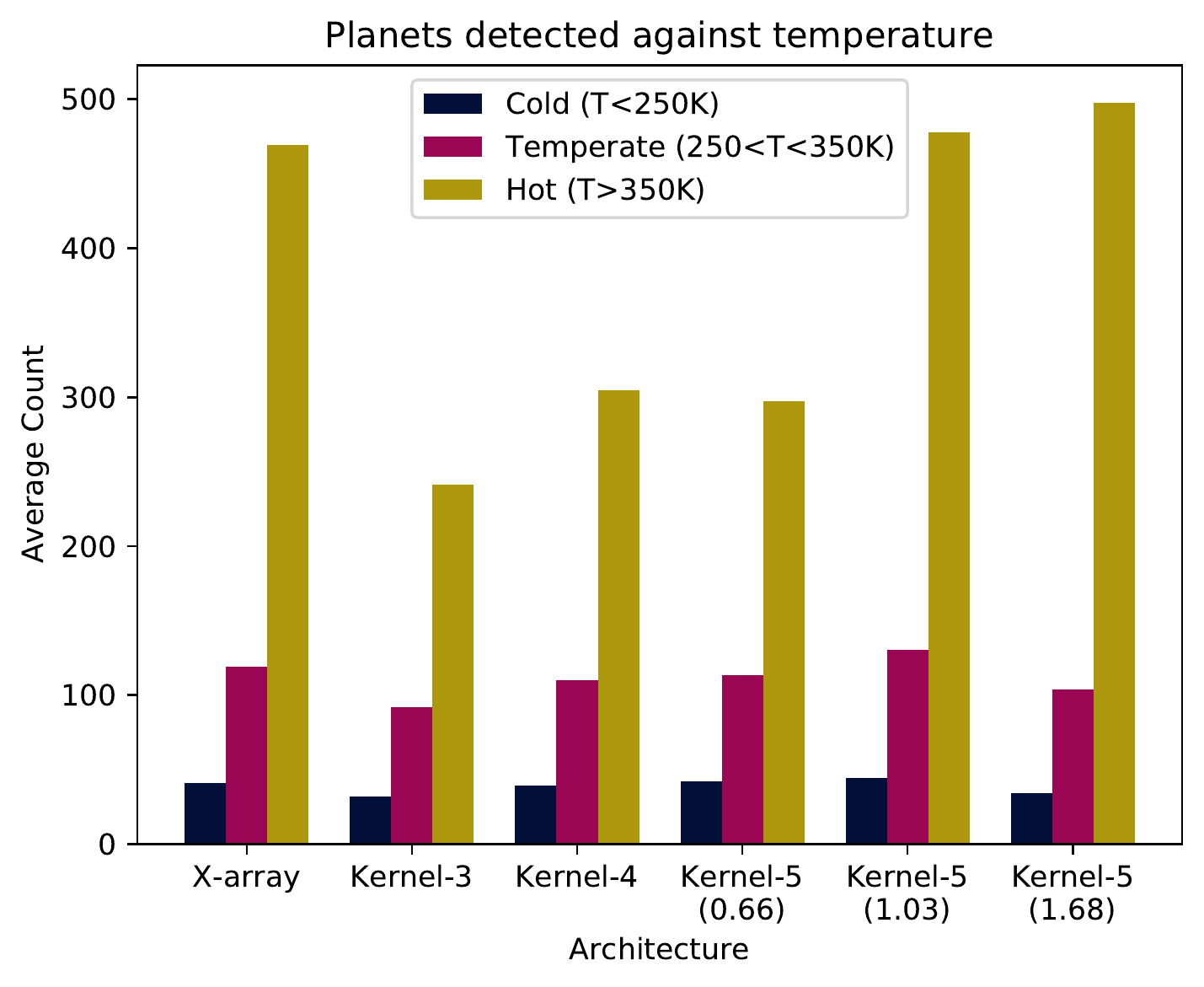}
      \caption{Number of planets detected for each nulling architecture, subdivided between hot, temperate and cold temperature bins, for a reference wavelength of 18~\textmu m.}
    \label{Img:Bar_temp}
\end{figure}

We also split the planets into temperature bins: cold planets at $<250$~K, temperate planets between 250~K and 350~K, and hot planets with $T>350$~K. Histograms of the detected count of planets is shown in Fig. \ref{Img:Bar_temp}. We see a pattern similar to the total exoplanets emerging for the numerous hot exoplanets, with the larger baseline Kernel-5 nullers and the X-array finding many more than the others. The other two subsets show only slight differences between configurations, although the Kernel-5 nuller with $\Gamma_B=1.03$ detects marginally more temperate exoplanets.

Finally, we show an amalgamation of Figs. \ref{Img:Bar_habitable},  \ref{Img:Bar_radius} and \ref{Img:Bar_temp} to investigate how each architecture responds to Earth twins: temperate, rocky planets in the habitable zone. This is shown in Fig. \ref{Img:Bar_habitable_rocky}. Here, we see that the architectures perform similarly, with all of them detecting between one and two Earth twins (again, assuming a 5~hr integration time). That being said, the Kernel-5 nuller at $\Gamma_B = 1.03$ has an 23\% detection increase over the X-array configuration with an average of 2.2 Earth twins detected.

\begin{figure}
  \begin{subfigure}{1\linewidth}
    \centering
    \includegraphics[width = 0.9\linewidth]{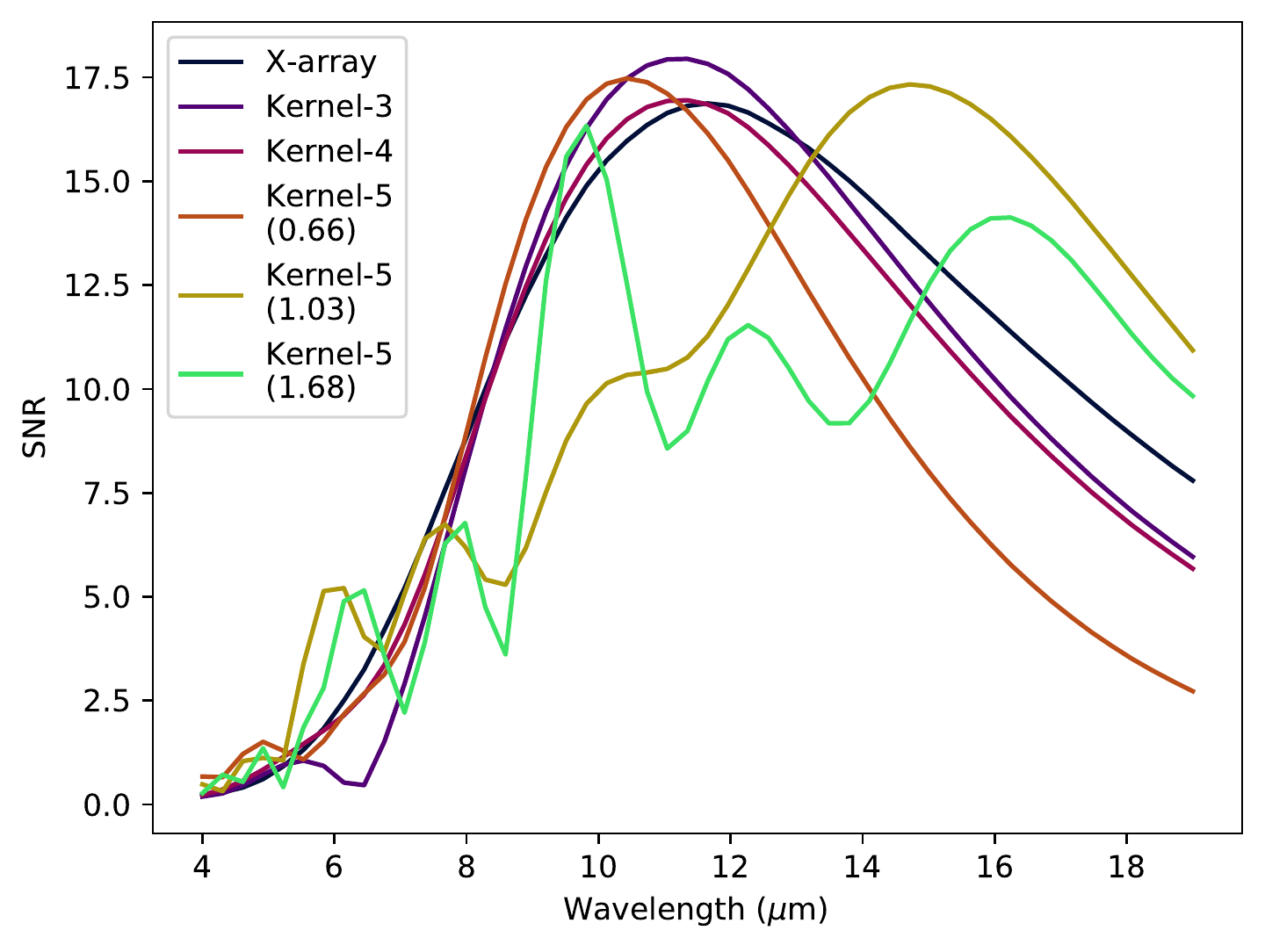}
    \caption{Super-Earth type planet around an M-dwarf.}
    \label{Img:SNR_wave_rocky}
  \end{subfigure}
  \hfill
  \begin{subfigure}{1\linewidth}
    \centering
    \includegraphics[width = 0.9\linewidth]{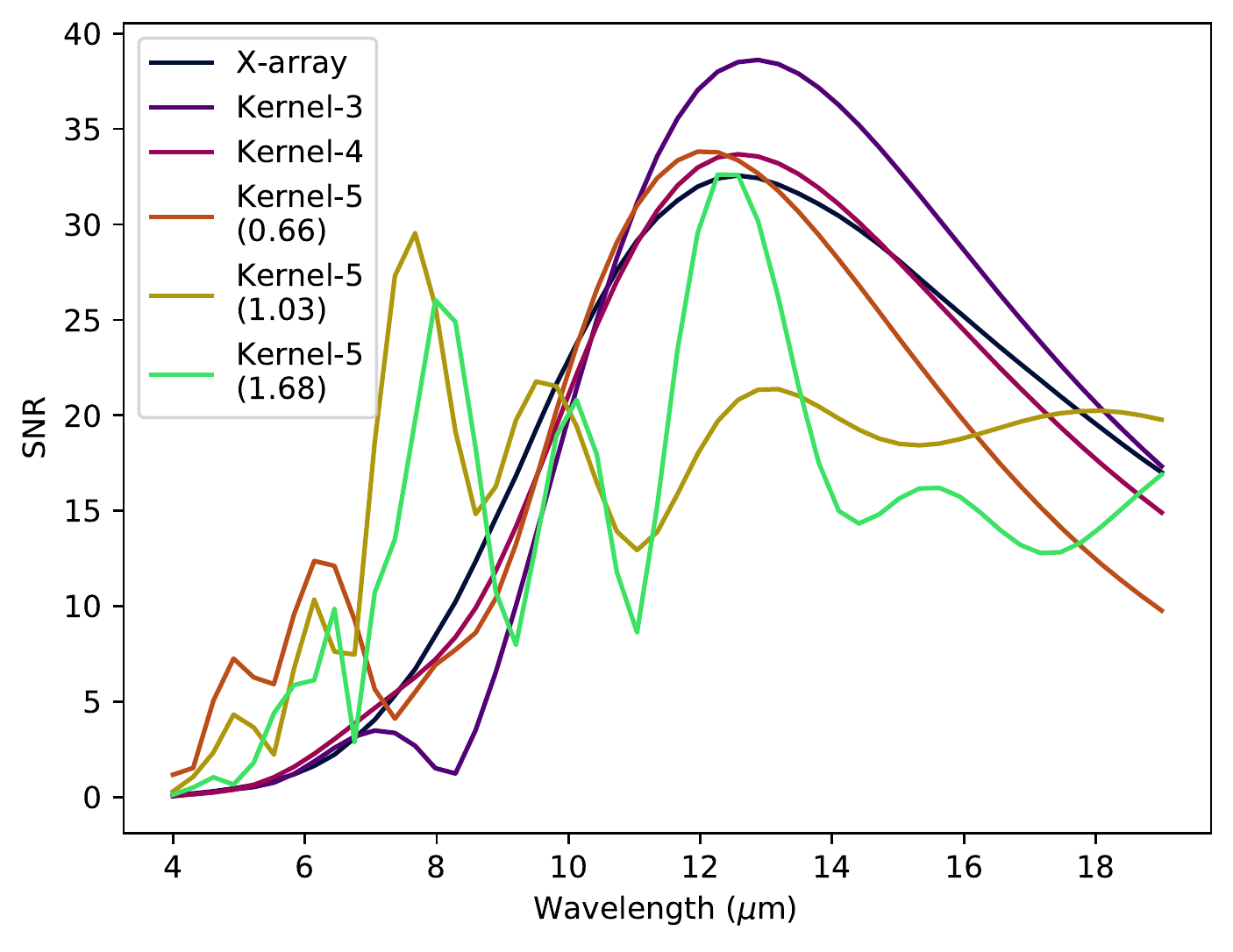}
    \caption{Neptune type planet around a K-dwarf.}
    \label{Img:SNR_wave_SuperEarth}
  \end{subfigure}
    \caption{SNR as a function of wavelength for the different architectures, taken for two example habitable zone planets.}
    \label{Img:SNR_wave}
\end{figure}

In Fig. \ref{Img:SNR_wave}, we plot the signal to noise ratio as a function of wavelength for two detectable planets in the habitable zone of their host stars; a rocky super-Earth around an M-dwarf ($D = 1.8~\text{pc}$, $T_p = 300~\text{K}$, $R_p = 1.3~R_\oplus$, $z=3.2~\text{zodis}$), and a gaseous Neptune-type planet around a K-dwarf ($D = 4.9~\text{pc}$, $T_p = 260~\text{K}$, $R_p = 3.9~R_\oplus$, $z=0.14~\text{zodis}$). The advantages of the Kernel-5 nuller are quite apparent: at smaller wavelengths, the SNR is higher than other architectures due to a deeper fourth-order null \citep{2013Guyon} reducing the stellar leakage (especially for the K dwarf), while the complex rotationally symmetric pattern allows for more transmission peaks as a function of wavelength than the other architectures. This results in a more consistent response as a function of wavelength and a higher overall signal to noise. We also note that, in the search phase, a random planet will have 50\% of its light going out the bright ports in the X-array configuration, while only losing 20\% of its light to the bright output for the Kernel-5 nuller.

Taken with the previous plots, it is quite apparent that when it comes to searching for exoplanets, particularly rocky, temperate planets in the habitable zone, the Kernel-5 nuller performs the best - especially when the baseline is scaled by $\Gamma_B = 1.03$. 

We briefly consider that, for detectable planets which have a maximum SNR at short wavelengths (4~\textmu m), the zodiacal light contribution is at a minimum; at most four orders of magnitude below the stellar leakage. As phase variation due to fringe tracking errors is linked to fluctuations in the null depth and by extension stellar leakage, we can estimate the RMS fringe tracking errors needed to be photon limited:
\begin{equation}
    \langle\phi^2\rangle F_\text{star} < \max{\left[\frac{P_\text{zodiacal}}{A}, F_\text{leakage}\right]}.
\end{equation}

At worst, therefore, we find that fringe tracking should be better than approximately 9~nm for an M-type star and 2.5nm for a G-type star in order to remain photon limited by zodiacal light, assuming an aperture diameter of 2~m. For a space based interferometer, this should be an achievable target.

\subsection{Characterisation Phase}

The other major component to a LIFE-type space interferometer mission is the characterisation phase - observing a known planet for a long enough period of time to receive a spectrum and possibly detect biosignatures (that is, `the presence of a gas or other feature that may be indicative of a biological agent' \citep{2018Schwieterman}). Hence the number of planets detectable is not the key parameter here, but rather which architectures produce a better SNR for a given amount of time. Conversely, the best architecture will provide quality spectra in shorter exposure times than the others, allowing for more targets to be observed. In the following discussion we keep the same basic setup from the search phase: a throughput of 5\%, five hour integration time, conservation of total collecting area and a reference wavelength of 18~\textmu m. The planet is now chosen to lie at the maximum of the transmission map (see Sect. \ref{Sec:planet_signal}).

\begin{figure}
    \centering
    \includegraphics[width = \linewidth]{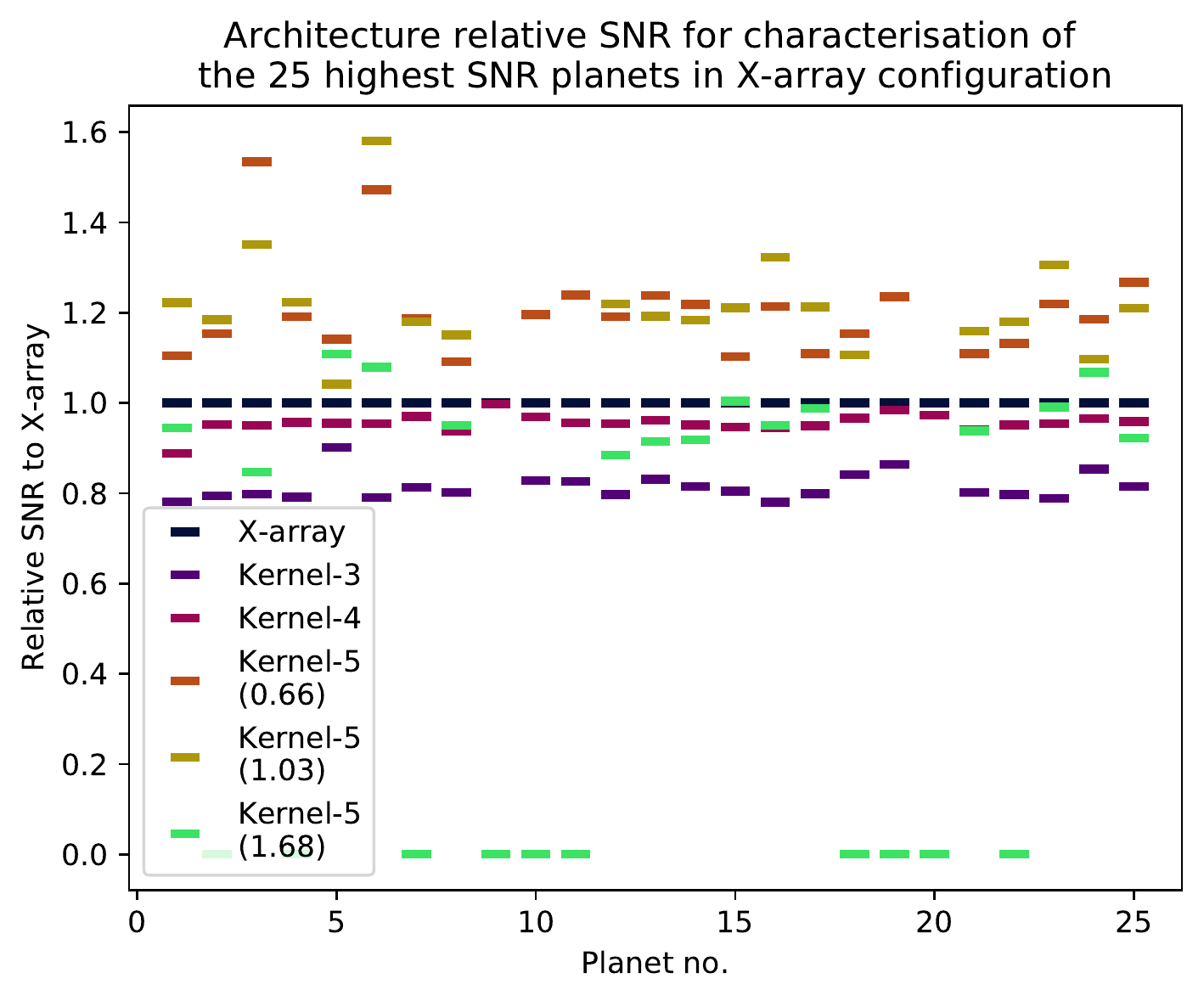}
    \caption{The SNRs for each architecture, relative to the X-array configuration, of the planets in the habitable zone with the 25 highest SNR when observing with the X-array design. }
    \label{Img:Char}
\end{figure}

In Fig. \ref{Img:Char}, we show the relative SNR of the 25 habitable zone planets with the highest signal in the X-array configuration for the six different architectures. Hence this plot should be inherently biased towards the X-array design. What we see instead is that the Kernel-5 nuller, particularly with $\Gamma_B = 0.66$ and 1.03, has a consistently higher SNR. Over the 25 planets, we find that these two Kernel-5 configurations on average achieve an SNR 1.2 times higher than the X-array, sometimes reaching as high as 1.6. Conversely, the Kernel-4 nuller performs marginally worse than the X-array (a similar result to the search phase) and the Kernel-3 nuller is one of the worst performing. We note that the points at zero relative SNR are caused by the configuration having a baseline outside of the 5 to 600~m range, and is particularly applicable to the Kernel-5 nullers with large baseline scale factors. 

\begin{figure}
  \begin{subfigure}{\linewidth}
    \centering
    \includegraphics[width = 0.9\linewidth]{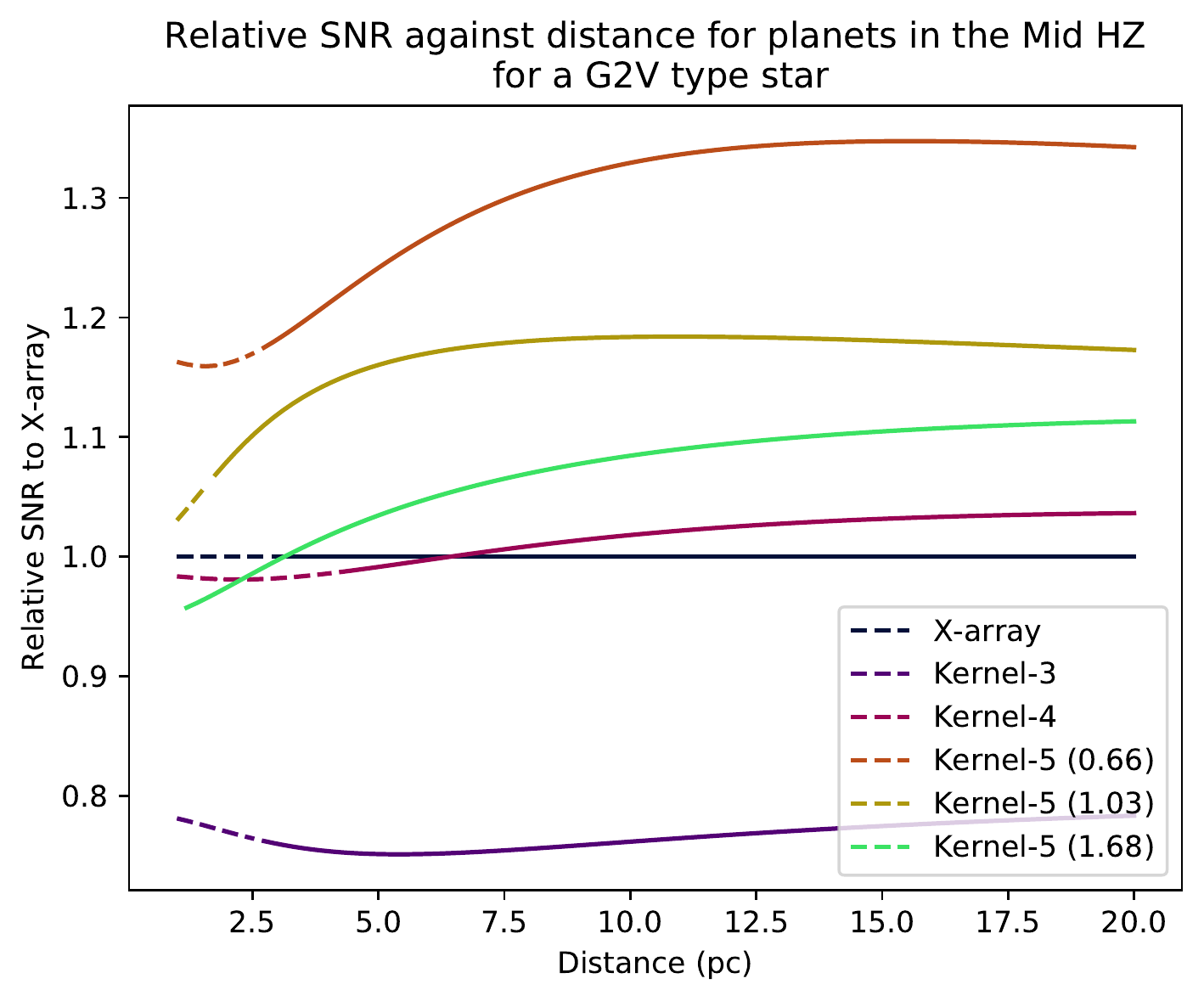}
    \caption{G2V type star based on the Sun}
  \end{subfigure}
    \begin{subfigure}{\linewidth}
    \centering
    \includegraphics[width = 0.9\linewidth]{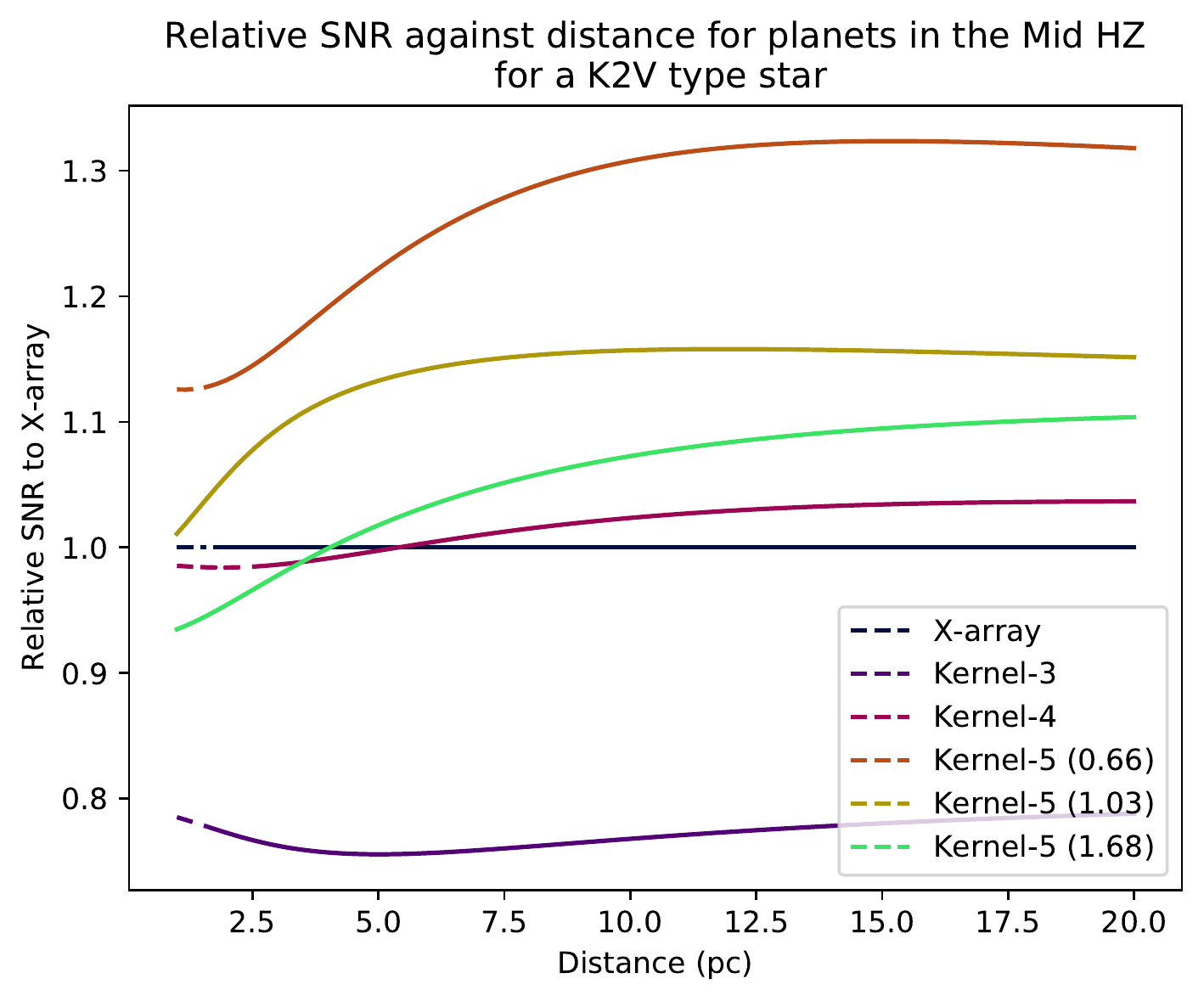}
    \caption{K2V type star based on Epsilon Eridani}
  \end{subfigure}
  \hfill
  \begin{subfigure}{\linewidth}
    \centering
    \includegraphics[width = 0.9\linewidth]{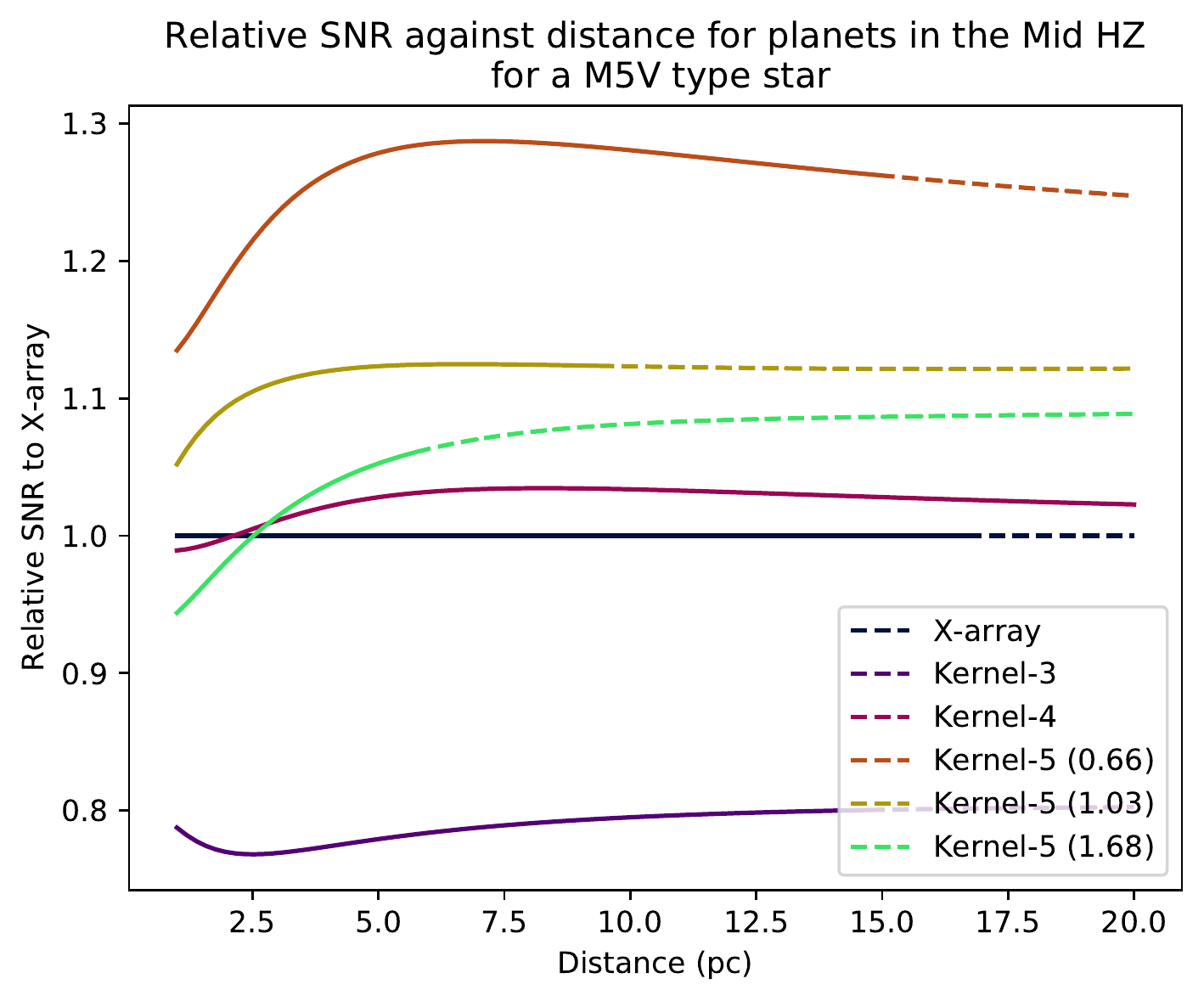}
    \caption{M5V type star based on Proxima Centauri}
  \end{subfigure}
    \caption{The SNRs for each architecture, relative to the X-array configuration, of a planet in the middle of the habitable zone of its host star as a function of stellar distance. Plotted for three stars with archetypal stellar types.}
    \label{Img:dist_plots}
\end{figure}

We also plot the relative SNR compared to the X-array design as a function of stellar distance of a planet in the middle of the habitable zone of stars of three difference spectral types. This is shown in Fig. \ref{Img:dist_plots}. Each simulated star had parameters based on an archetypal star of that spectral type: a G2V star based on the Sun, a K2V star based on Epsilon Eridani, and an M5V star based on Proxima Centauri. The dashed lines in the plot indicate when the configuration's baseline moves outside of the acceptable range. 

It is evident that these plots echo that of Fig. \ref{Img:Char}, in that the Kernel-5 nuller produces a better SNR than the X-array consistently. Interestingly though, the Kernel-5 variant with $\Gamma_B = 0.66$ is significantly better than $\Gamma_B = 1.03$ at all distances, and especially with M type stars where the baseline is still short enough to probe stars as far out as 15~pc. It is also better quantitatively, exhibiting 1.3 times the SNR of the X-array design on average, though this falters for stars closer than 5~pc.  Based on this, it seems that the $\Gamma_B = 0.66$ Kernel-5 nuller is the best architecture for characterisation by a substantial margin. 

One potential reason for this is the `zoomed-in' nature of the transmission map for a smaller scale factor; this produces a greater SNR across the wavelength-dependent angular positions when the planet is at the point of maximum transmission when compared to a `zoomed-out' map with larger radial variance. This same feature is a disadvantage in search mode: if the planet is in a transmission minima, it may be in the same minima at more wavelength-dependent positions compared to an architecture with a larger scale factor. Nevertheless, as $\Gamma_B$ is simply a baseline scaling factor, it would be trivial to implement the control such that a Kernel-5 nuller uses a longer $\Gamma_B=1.03$ during search phase and a shorter $\Gamma_B=0.66$ during characterisation. 

\red{It is also worth noting that the Kernel-5 nuller's two kernel maps combined contain more spatial information than the single chopped output of the X-array. This would allow the spectrum extraction algorithm (such as the one posited by \cite{LIFE2}), in the almost certain case of a multi-planetary system, to be able to more easily distinguish between the fluxes of multiple planets and increase the constraints of the exoplanet's astrophysical properties.}

Finally, we comment here that we also performed a similar analysis with seven telescopes arranged in a heptagonal configuration. We found that it performed $\sim$15\% worse in detection than the Kernel-5 nuller, but $\sim$15\% better in characterisation, likely due to the presence of a sixth order null. However, due to the great increase in complexity, both in terms of optical design as well as management of spacecraft; and the meagre improvement, we did not pursue this line of research further. We hypothesise that the characterisation will continue to improve by increasing the number of telescopes (particularly with odd numbers in order to have a simpler optical design and access to higher order nulls) up to a point where stellar leakage is no longer dominant in the wavelength range. Despite this, the additional design complexity will likely counteract this advantage.

\section{Conclusions}

With optical space interferometry once again holding potential as a future mission (in the form of LIFE), it is vital to take a critical look at the assumptions made in the past and adapt them to work in the future. Through our simulations, we have found that through the use of a kernel-nulling beam combiner and when conserving total collecting area, an architecture consisting of five telescopes in a regular pentagonal configuration provides better scientific return than the X-array design inherited from the \textit{Darwin} and TPF-I trade studies. This holds true for both search and characterisation:
\begin{itemize}
    \item In search mode, with a reference wavelength of $\lambda_B=$18~\textmu m, the Kernel-5 nuller with a baseline scale factor of $\Gamma_B = 1.03$ would detect 23\% more Earth twins (temperate, rocky and in the habitable zone) than the X-array. It also finds considerably more rocky planets, habitable zone planets and total planets than its other architecture counterparts. 
    \item In characterisation mode, again with a reference wavelength of $\lambda_B=$18~\textmu m, the Kernel-5 nuller with a baseline scale factor of $\Gamma_B = 0.66$ has on average an SNR between 1.2 and 1.3 times greater than the X-array. This holds for planets around GKM stars, as well as at a majority of stellar distances.
    \item The fact that search and characterisation modes favour different baseline scale lengths is not a problem in this study, as this scaling factor can be changed in real time depending on what mode the interferometer is undertaking. 
\end{itemize}

Hence, we recommend that future studies and simulations based around a large, exoplanet hunting, optical space interferometry mission, such as LIFE, consider adopting a Kernel-5 nulling architecture as the basis of the design.  There may also be further benefits to this architecture involving redundancy; that the failure of one of the five collecting spacecraft may not result in the failure of the mission as a whole. Both this and realistic instrumental noise will be addressed in a follow up article \red{\citep{LIFE7}}.

We note here two small caveats to the recommendation of the Kernel-5 architecture. The first is the additional complexity of having one more spacecraft, though this may be be of benefit due to the added redundancy of an extra telescope. Second, and arguably more importantly, to implement the Kernel-5 nuller requires a range of achromatic phase shifts that deviate from the standard $\pi$ and $\pi/2$ phase shifts used in the X-array. A potential implementation of such a beam combination scheme in bulk optics will also be discussed in a follow up article \red{\citep{LIFE7}}. In principle, photonics may be able to provide an arbitrary achromatic phase shift, but this needs to yet be successfully demonstrated at 10 microns and under cryogenic conditions. 

The resurrection of optical space interferometry as a tool for exoplanet science holds extreme potential in revolutionising the field and providing humanity with the possible first signs of life on another world. Simultaneously looking back at the past of \textit{Darwin} and TPF-I, learning from both the achievements and failures made in that era, while also looking forward at future technologies and applying new research collaboratively, is likely the only way that this dream from decades ago may one day see the faint light of planets far, far away.

\section*{Acknowledgements}

This project has made use of P-Pop. 
This research was supported by the ANU Futures scheme and by the Australian Government through the Australian Research Council's Discovery Projects funding scheme (project DP200102383). 
The JWST Background program was written by Jane Rigby (GSFC, Jane.Rigby@nasa.gov) and Klaus Pontoppidan (STScI, pontoppi@stsci.edu). The associated background cache was prepared by Wayne Kinzel at STScI, and is the same as used by the JWST Exposure Time Calculator.
We also greatly thank Jens Kammerer, Sascha Quanz and the rest of the LIFE team for helpful discussions.

The code used to generate these simulations can be found publicly at \url{https://github.com/JonahHansen/LifeTechSim}. 




\bibliographystyle{aa}
\bibliography{bibliography} 

\begin{thebibliography}{41}
\expandafter\ifx\csname natexlab\endcsname\relax\def\natexlab#1{#1}\fi

\bibitem[{{Angel} \& {Woolf}(1997)}]{1997Angel}
{Angel}, J.~R.~P. \& {Woolf}, N.~J. 1997, \apj, 475, 373

\bibitem[{{Berger} {et~al.}(2020){Berger}, {Huber}, {Gaidos}, {van Saders}, \&
  {Weiss}}]{2020Berger}
{Berger}, T.~A., {Huber}, D., {Gaidos}, E., {van Saders}, J.~L., \& {Weiss},
  L.~M. 2020, \aj, 160, 108

\bibitem[{{Borucki} {et~al.}(2010){Borucki}, {Koch}, {Basri}, {Batalha},
  {Brown}, {Caldwell}, {Caldwell}, {Christensen-Dalsgaard}, {Cochran},
  {DeVore}, {Dunham}, {Dupree}, {Gautier}, {Geary}, {Gilliland}, {Gould},
  {Howell}, {Jenkins}, {Kondo}, {Latham}, {Marcy}, {Meibom}, {Kjeldsen},
  {Lissauer}, {Monet}, {Morrison}, {Sasselov}, {Tarter}, {Boss}, {Brownlee},
  {Owen}, {Buzasi}, {Charbonneau}, {Doyle}, {Fortney}, {Ford}, {Holman},
  {Seager}, {Steffen}, {Welsh}, {Rowe}, {Anderson}, {Buchhave}, {Ciardi},
  {Walkowicz}, {Sherry}, {Horch}, {Isaacson}, {Everett}, {Fischer}, {Torres},
  {Johnson}, {Endl}, {MacQueen}, {Bryson}, {Dotson}, {Haas}, {Kolodziejczak},
  {Van Cleve}, {Chandrasekaran}, {Twicken}, {Quintana}, {Clarke}, {Allen},
  {Li}, {Wu}, {Tenenbaum}, {Verner}, {Bruhweiler}, {Barnes}, \&
  {Prsa}}]{Kepler}
{Borucki}, W.~J., {Koch}, D., {Basri}, G., {et~al.} 2010, Science, 327, 977

\bibitem[{{Bracewell}(1978)}]{1978Bracewell}
{Bracewell}, R.~N. 1978, \nat, 274, 780

\bibitem[{{Broeg} {et~al.}(2013){Broeg}, {Fortier}, {Ehrenreich}, {Alibert},
  {Baumjohann}, {Benz}, {Deleuil}, {Gillon}, {Ivanov}, {Liseau}, {Meyer},
  {Oloffson}, {Pagano}, {Piotto}, {Pollacco}, {Queloz}, {Ragazzoni}, {Renotte},
  {Steller}, \& {Thomas}}]{CHEOPS}
{Broeg}, C., {Fortier}, A., {Ehrenreich}, D., {et~al.} 2013, in European
  Physical Journal Web of Conferences, Vol.~47, European Physical Journal Web
  of Conferences, 03005

\bibitem[{{Cabrera} {et~al.}(2020){Cabrera}, {McMurtry}, {Forrest}, {Pipher},
  {Dorn}, \& {Lee}}]{2020Cabrera}
{Cabrera}, M.~S., {McMurtry}, C.~W., {Forrest}, W.~J., {et~al.} 2020, Journal
  of Astronomical Telescopes, Instruments, and Systems, 6, 011004

\bibitem[{{Chen} \& {Kipping}(2017)}]{2017Chen}
{Chen}, J. \& {Kipping}, D. 2017, \apj, 834, 17

\bibitem[{{Claret} \& {Bloemen}(2011)}]{2011Claret}
{Claret}, A. \& {Bloemen}, S. 2011, \aap, 529, A75

\bibitem[{{Dannert} {et~al.}(2022){Dannert}, {Ottiger}, {Quanz}, {Laugier},
  {Fontanet}, {Gheorghe}, {Absil}, {Dandumont}, {Defr{\`e}re}, {Gasc{\'o}n},
  {Glauser}, {Kammerer}, {Lichtenberg}, {Linz}, {Loicq}, \& {the LIFE
  collaboration}}]{LIFE2}
{Dannert}, F., {Ottiger}, M., {Quanz}, S.~P., {et~al.} 2022, \aap, in press

\bibitem[{{Defr{\`e}re} {et~al.}(2010){Defr{\`e}re}, {Absil}, {den Hartog},
  {Hanot}, \& {Stark}}]{2010Defrere}
{Defr{\`e}re}, D., {Absil}, O., {den Hartog}, R., {Hanot}, C., \& {Stark}, C.
  2010, \aap, 509, A9

\bibitem[{{Ertel} {et~al.}(2020){Ertel}, {Defr{\`e}re}, {Hinz}, {Mennesson},
  {Kennedy}, {Danchi}, {Gelino}, {Hill}, {Hoffmann}, {Mazoyer}, {Rieke},
  {Shannon}, {Stapelfeldt}, {Spalding}, {Stone}, {Vaz}, {Weinberger},
  {Willems}, {Absil}, {Arbo}, {Bailey}, {Beichman}, {Bryden}, {Downey},
  {Durney}, {Esposito}, {Gaspar}, {Grenz}, {Haniff}, {Leisenring}, {Marion},
  {McMahon}, {Millan-Gabet}, {Montoya}, {Morzinski}, {Perera}, {Pinna}, {Pott},
  {Power}, {Puglisi}, {Roberge}, {Serabyn}, {Skemer}, {Su}, {Vaitheeswaran}, \&
  {Wyatt}}]{2020Ertel}
{Ertel}, S., {Defr{\`e}re}, D., {Hinz}, P., {et~al.} 2020, \aj, 159, 177

\bibitem[{{Fulton} \& {Petigura}(2018)}]{2018Fulton}
{Fulton}, B.~J. \& {Petigura}, E.~A. 2018, \aj, 156, 264

\bibitem[{{Gheorghe} {et~al.}(2020){Gheorghe}, {Glauser}, {Ergenzinger},
  {Quanz}, {Defr{\`e}re}, \& {Kuhn}}]{2020Gheorghe}
{Gheorghe}, A.~A., {Glauser}, A.~M., {Ergenzinger}, K., {et~al.} 2020, in
  Society of Photo-Optical Instrumentation Engineers (SPIE) Conference Series,
  Vol. 11446, Society of Photo-Optical Instrumentation Engineers (SPIE)
  Conference Series, 114462N

\bibitem[{{Gretzinger} {et~al.}(2019){Gretzinger}, {Gross}, {Arriola}, \&
  {Withford}}]{2019Gretzinger}
{Gretzinger}, T., {Gross}, S., {Arriola}, A., \& {Withford}, M.~J. 2019, Optics
  Express, 27, 8626

\bibitem[{{Guyon} {et~al.}(2013){Guyon}, {Mennesson}, {Serabyn}, \&
  {Martin}}]{2013Guyon}
{Guyon}, O., {Mennesson}, B., {Serabyn}, E., \& {Martin}, S. 2013, \pasp, 125,
  951

\bibitem[{{Hansen} {et~al.}(2021){Hansen}, {Casagrande}, {Ireland}, \&
  {Lin}}]{2021HansenPlanets}
{Hansen}, J.~T., {Casagrande}, L., {Ireland}, M.~J., \& {Lin}, J. 2021, \mnras,
  501, 5309

\bibitem[{{Hansen} \& {Ireland}(2020)}]{2020HansenHonours}
{Hansen}, J.~T. \& {Ireland}, M.~J. 2020, \pasa, 37, e019

\bibitem[{{Hansen} {et~al.}(2022){Hansen}, {Ireland}, {Laugier}, \& {the LIFE
  collaboration}}]{LIFE7}
{Hansen}, J.~T., {Ireland}, M.~J., {Laugier}, R., \& {the LIFE collaboration}.
  2022, arXiv e-prints, arXiv:2204.12291

\bibitem[{{Kammerer} \& {Quanz}(2018)}]{2018Kammerer}
{Kammerer}, J. \& {Quanz}, S.~P. 2018, \aap, 609, A4

\bibitem[{{Karlsson} {et~al.}(2004){Karlsson}, {Wallner}, {Perdigues Armengol},
  \& {Absil}}]{2004Karlsson}
{Karlsson}, A.~L., {Wallner}, O., {Perdigues Armengol}, J.~M., \& {Absil}, O.
  2004, in Society of Photo-Optical Instrumentation Engineers (SPIE) Conference
  Series, Vol. 5491, New Frontiers in Stellar Interferometry, ed. W.~A.
  {Traub}, 831

\bibitem[{{Kenchington Goldsmith} {et~al.}(2017){Kenchington Goldsmith},
  {Cvetojevic}, {Ireland}, \& {Madden}}]{2017Harry}
{Kenchington Goldsmith}, H.-D., {Cvetojevic}, N., {Ireland}, M., \& {Madden},
  S. 2017, Optics Express, 25, 3038

\bibitem[{{Kennedy} {et~al.}(2015){Kennedy}, {Wyatt}, {Bailey}, {Bryden},
  {Danchi}, {Defr{\`e}re}, {Haniff}, {Hinz}, {Lebreton}, {Mennesson},
  {Millan-Gabet}, {Morales}, {Pani{\'c}}, {Rieke}, {Roberge}, {Serabyn},
  {Shannon}, {Skemer}, {Stapelfeldt}, {Su}, \& {Weinberger}}]{2015Kennedy}
{Kennedy}, G.~M., {Wyatt}, M.~C., {Bailey}, V., {et~al.} 2015, \apjs, 216, 23

\bibitem[{{Kopparapu} {et~al.}(2018){Kopparapu}, {H{\'e}brard}, {Belikov},
  {Batalha}, {Mulders}, {Stark}, {Teal}, {Domagal-Goldman}, \&
  {Mandell}}]{2018Kopparapu}
{Kopparapu}, R.~K., {H{\'e}brard}, E., {Belikov}, R., {et~al.} 2018, \apj, 856,
  122

\bibitem[{{Kopparapu} {et~al.}(2013){Kopparapu}, {Ramirez}, {Kasting}, {Eymet},
  {Robinson}, {Mahadevan}, {Terrien}, {Domagal-Goldman}, {Meadows}, \&
  {Deshpande}}]{2013Kopparapu}
{Kopparapu}, R.~K., {Ramirez}, R., {Kasting}, J.~F., {et~al.} 2013, \apj, 765,
  131

\bibitem[{{Laugier} {et~al.}(2020){Laugier}, {Cvetojevic}, \&
  {Martinache}}]{2020Laugier}
{Laugier}, R., {Cvetojevic}, N., \& {Martinache}, F. 2020, \aap, 642, A202

\bibitem[{{Lay}(2004)}]{2004LayNoise}
{Lay}, O.~P. 2004, \ao, 43, 6100

\bibitem[{{Lay}(2006)}]{2006Lay}
{Lay}, O.~P. 2006, in Society of Photo-Optical Instrumentation Engineers (SPIE)
  Conference Series, Vol. 6268, Society of Photo-Optical Instrumentation
  Engineers (SPIE) Conference Series, ed. J.~D. {Monnier}, M.~{Sch{\"o}ller},
  \& W.~C. {Danchi}, 62681A

\bibitem[{{Lay} {et~al.}(2005){Lay}, {Gunter}, {Hamlin}, {Henry}, {Li},
  {Martin}, {Purcell}, {Ware}, {Wertz}, \& {Noecker}}]{2005LayXarray}
{Lay}, O.~P., {Gunter}, S.~M., {Hamlin}, L.~A., {et~al.} 2005, in Society of
  Photo-Optical Instrumentation Engineers (SPIE) Conference Series, Vol. 5905,
  Techniques and Instrumentation for Detection of Exoplanets II, ed. D.~R.
  {Coulter}, 8--20

\bibitem[{{L{\'e}ger} {et~al.}(1995){L{\'e}ger}, {Puget}, {Mariotti}, {Rouan},
  \& {Schneider}}]{1995Leger}
{L{\'e}ger}, A., {Puget}, J.~J., {Mariotti}, J.~M., {Rouan}, D., \&
  {Schneider}, J. 1995, \ssr, 74, 163

\bibitem[{Loreggia {et~al.}(2018)Loreggia, Fineschi, Bemporad, Capobianco,
  Nicolini, Zangrilli, Casti, Landini, Baccani, Romoli, Buckley, Thizy, Denis,
  Ledent, Marquet, Galano, Belluso, Accatino, Terenzi, Morgante, Riva,
  Moschetti, Calderoni, Pieraccini, \& Noce}]{2018Loreggia}
Loreggia, D., Fineschi, S., Bemporad, A., {et~al.} 2018, in Optical Instrument
  Science, Technology, and Applications, 1069503, ed. R.~N. Youngworth \&
  N.~Haverkamp, Vol. 10695 (Frankfurt, Germany: SPIE)

\bibitem[{{Martinache} \& {Ireland}(2018)}]{2018Martinache}
{Martinache}, F. \& {Ireland}, M.~J. 2018, \aap, 619, A87

\bibitem[{{Mayor} {et~al.}(2003){Mayor}, {Pepe}, {Queloz}, {Bouchy},
  {Rupprecht}, {Lo Curto}, {Avila}, {Benz}, {Bertaux}, {Bonfils}, {Dall},
  {Dekker}, {Delabre}, {Eckert}, {Fleury}, {Gilliotte}, {Gojak}, {Guzman},
  {Kohler}, {Lizon}, {Longinotti}, {Lovis}, {Megevand}, {Pasquini}, {Reyes},
  {Sivan}, {Sosnowska}, {Soto}, {Udry}, {van Kesteren}, {Weber}, \&
  {Weilenmann}}]{HARPS}
{Mayor}, M., {Pepe}, F., {Queloz}, D., {et~al.} 2003, The Messenger, 114, 20

\bibitem[{{Petigura} {et~al.}(2018){Petigura}, {Marcy}, {Winn}, {Weiss},
  {Fulton}, {Howard}, {Sinukoff}, {Isaacson}, {Morton}, \&
  {Johnson}}]{2018Petigura}
{Petigura}, E.~A., {Marcy}, G.~W., {Winn}, J.~N., {et~al.} 2018, \aj, 155, 89

\bibitem[{{Quanz} {et~al.}(2022){Quanz}, {Ottiger}, {Fontanet}, {Kammerer},
  {Menti}, {Dannert}, {Gheorghe}, {Absil}, {Airapetian}, {Alei}, {Allart},
  {Angerhausen}, {Blumenthal}, {Buchhave}, {Cabrera},
  {Carri{\'o}n-Gonz{\'a}lez}, {Chauvin}, {Danchi}, {Dandumont}, {Defr{\`e}re},
  {Dorn}, {Ehrenreich}, {Ertel}, {Fridlund}, {Garc{\'\i}a Mu{\~n}oz},
  {Gasc{\'o}n}, {Girard}, {Glauser}, {Grenfell}, {Guidi}, {Hagelberg},
  {Helled}, {Ireland}, {Kopparapu}, {Korth}, {Kozakis}, {Kraus}, {L{\'e}ger},
  {Leedj{\"a}rv}, {Lichtenberg}, {Lillo-Box}, {Linz}, {Liseau}, {Loicq},
  {Mahendra}, {Malbet}, {Mathew}, {Mennesson}, {Meyer}, {Mishra},
  {Molaverdikhani}, {Noack}, {Oza}, {Pall{\'e}}, {Parviainen}, {Quirrenbach},
  {Rauer}, {Ribas}, {Rice}, {Romagnolo}, {Rugheimer}, {Schwieterman},
  {Serabyn}, {Sharma}, {Stassun}, {Szul{\'a}gyi}, {Wang}, {Wunderlich},
  {Wyatt}, \& {the LIFE collaboration}}]{LIFEPaper1}
{Quanz}, S.~P., {Ottiger}, M., {Fontanet}, E., {et~al.} 2022, \aap, in press

\bibitem[{{Ricker} {et~al.}(2015){Ricker}, {Winn}, {Vanderspek}, {Latham},
  {Bakos}, {Bean}, {Berta-Thompson}, {Brown}, {Buchhave}, {Butler}, {Butler},
  {Chaplin}, {Charbonneau}, {Christensen-Dalsgaard}, {Clampin}, {Deming},
  {Doty}, {De Lee}, {Dressing}, {Dunham}, {Endl}, {Fressin}, {Ge}, {Henning},
  {Holman}, {Howard}, {Ida}, {Jenkins}, {Jernigan}, {Johnson}, {Kaltenegger},
  {Kawai}, {Kjeldsen}, {Laughlin}, {Levine}, {Lin}, {Lissauer}, {MacQueen},
  {Marcy}, {McCullough}, {Morton}, {Narita}, {Paegert}, {Palle}, {Pepe},
  {Pepper}, {Quirrenbach}, {Rinehart}, {Sasselov}, {Sato}, {Seager},
  {Sozzetti}, {Stassun}, {Sullivan}, {Szentgyorgyi}, {Torres}, {Udry}, \&
  {Villasenor}}]{TESS}
{Ricker}, G.~R., {Winn}, J.~N., {Vanderspek}, R., {et~al.} 2015, Journal of
  Astronomical Telescopes, Instruments, and Systems, 1, 014003

\bibitem[{{Rogers}(2015)}]{2015Rogers}
{Rogers}, L.~A. 2015, \apj, 801, 41

\bibitem[{{Schwieterman} {et~al.}(2018){Schwieterman}, {Kiang}, {Parenteau},
  {Harman}, {DasSarma}, {Fisher}, {Arney}, {Hartnett}, {Reinhard}, {Olson},
  {Meadows}, {Cockell}, {Walker}, {Grenfell}, {Hegde}, {Rugheimer}, {Hu}, \&
  {Lyons}}]{2018Schwieterman}
{Schwieterman}, E.~W., {Kiang}, N.~Y., {Parenteau}, M.~N., {et~al.} 2018,
  Astrobiology, 18, 663

\bibitem[{{Stenborg} {et~al.}(2021){Stenborg}, {Howard}, {Hess}, \&
  {Gallagher}}]{2021Stenborg}
{Stenborg}, G., {Howard}, R.~A., {Hess}, P., \& {Gallagher}, B. 2021, \aap,
  650, A28

\bibitem[{{Velusamy} {et~al.}(2003){Velusamy}, {Angel}, {Eatchel}, {Tenerelli},
  \& {Woolf}}]{2003Velusamy}
{Velusamy}, T., {Angel}, R.~P., {Eatchel}, A., {Tenerelli}, D., \& {Woolf},
  N.~J. 2003, in ESA Special Publication, Vol. 539, Earths: DARWIN/TPF and the
  Search for Extrasolar Terrestrial Planets, ed. M.~{Fridlund}, T.~{Henning},
  \& H.~{Lacoste}, 631--636

\bibitem[{{Vogt} {et~al.}(1994){Vogt}, {Allen}, {Bigelow}, {Bresee}, {Brown},
  {Cantrall}, {Conrad}, {Couture}, {Delaney}, {Epps}, {Hilyard}, {Hilyard},
  {Horn}, {Jern}, {Kanto}, {Keane}, {Kibrick}, {Lewis}, {Osborne},
  {Pardeilhan}, {Pfister}, {Ricketts}, {Robinson}, {Stover}, {Tucker}, {Ward},
  \& {Wei}}]{HIRES}
{Vogt}, S.~S., {Allen}, S.~L., {Bigelow}, B.~C., {et~al.} 1994, in Society of
  Photo-Optical Instrumentation Engineers (SPIE) Conference Series, Vol. 2198,
  Instrumentation in Astronomy VIII, ed. D.~L. {Crawford} \& E.~R. {Craine},
  362

\bibitem[{{Voyage 2050 Senior Committee}(2021)}]{2021ESAVoyage}
{Voyage 2050 Senior Committee}. 2021, Voyage 2050 - Final Recommendations from
  the Voyage 2050 Senior Committee

\end{thebibliography}







\label{lastpage}
\end{document}